\documentclass[useAMS,usenatbib,usegraphicx]{mn2e}

\newcommand{\Ni}{$^{56}$Ni}

\newcommand{\Msun}{$M_\odot$}

\title[Supernovae from RSGs with Extensive Mass Loss]
{
Supernovae from Red Supergiants with Extensive Mass Loss
}
\author[T. Moriya et al.]
{Takashi Moriya$^{1,2,3}$\thanks{takashi.moriya@ipmu.jp},
Nozomu Tominaga$^{4,1}$,
Sergei I. Blinnikov$^{5,6,7,1}$,  \newauthor
Petr V. Baklanov$^{5}$, \&
Elena I. Sorokina$^{6,7}$ \\
$^{1}$
Institute for the Physics and Mathematics of the Universe,
Todai Institutes for Advanced Study,
University of Tokyo, \\ Kashiwanoha 5-1-5, Kashiwa, Chiba 277-8583, Japan
\\
$^{2}$
Department of Astronomy, Graduate School of Science, University of
Tokyo, 7-3-1 Hongo, Bunkyo-ku, Tokyo 113-0033, Japan \\
$^{3}$
Research Center for the Early Universe, Graduate School of Science, University of Tokyo,
7-3-1 Hongo, Bunkyo-ku, Tokyo 113-0033, Japan\\
$^{4}$
Department of Physics, Faculty of Science and Engineering, Konan University, 8-9-1 Okamoto, Kobe, Hyogo 658-8501, Japan\\
$^{5}$
Institute for Theoretical and Experimental Physics, Bolshaya Cheremushkinskaya 25, 117218 Moscow, Russia\\ 
$^{6}$
Sternberg Astronomical Institute, Moscow University, Universitetski pr. 13, 119992 Moscow, Russia\\
$^{7}$
Max-Planck-Institut f\"ur Astrophysik, Karl-Schwarzschild-Str. 1, 85741 Garching, Germany
}
\begin{document}


\pagerange{\pageref{firstpage}--\pageref{lastpage}} \pubyear{2011}

\maketitle

\label{firstpage}

\begin{abstract}
We calculate multicolor light curves (LCs) of supernovae (SNe) from red supergiants (RSGs)
exploded within dense circumstellar medium (CSM).
Multicolor LCs are calculated by using a multi-group
radiation hydrodynamics code \verb|STELLA|.
If CSM is dense enough,
the shock breakout signal is delayed and smeared by CSM and
kinetic energy of SN ejecta
is efficiently converted to thermal energy which is eventually released as radiation. 
We find that explosions of RSGs are affected by CSM in early epochs
when mass-loss rate just before the explosions
is higher than $\sim 10^{-4}~M_\odot~\mathrm{yr^{-1}}$.
Their characteristic features are
that the LC has a luminous round peak followed by a flat LC,
that multicolor LCs are simultaneously bright in ultraviolet and optical at the peak,
and that photospheric velocity is very low at these epochs.
We calculate LCs for various CSM conditions and explosion properties, i.e.,
mass-loss rates, radii of CSM, density slopes of CSM, explosion
energies of SN ejecta, and SN progenitors inside, 
to see their influence on LCs.
We compare our model LCs to those of ultraviolet-bright Type IIP SN
2009kf and show that the mass-loss rate of the progenitor
of SN 2009kf just before the explosion 
is likely to be higher than $10^{-4}~M_\odot~\mathrm{yr^{-1}}$.
Combined with the fact that SN 2009kf is likely to be an energetic explosion
and has large \Ni~production,
which implies that the progenitor of SN 2009kf is a massive
RSG, our results indicate that there could be some mechanism
to induce extensive mass loss in massive RSGs
just before their explosions.
\end{abstract}


\section{Introduction}
What kind of stars becomes which kind of supernovae (SNe)?
There still no conclusive answer to this simple question.
Looking into the case of core-collapse SNe (CCSNe) which result
from death of massive stars, there are several discrepancies between answers
from observers and theorists.
Observers get information about the progenitors of CCSNe by
directly getting images of the SN progenitors
by looking into the archival images of the SN site
\citep[][and the references therein]{smarttall}
or statistically analysing the number of SNe appeared
\citep[e.g.,][]{loss,boissier,ptf}.
On the other hand, theorists have been calculating
evolution of massive stars for decades to get progenitor models of SNe.
Theoretical pictures of single star evolution without rotation
are well summarized in \citet{heger}.

One of the big discrepancies raised recently between observers and
theorists is the zero-age main-sequence (ZAMS) mass range of Type IIP SN progenitors.
Type IIP SNe are the most common type of CCSNe which results
from explosions of red supergiants (RSGs) with a large amount of
hydrogen\footnote{For the SN classification, see \citet{filippenko}
for a review.}.
Both observers and theorists suggest that
the minimum ZAMS mass of Type IIP SN progenitors
at solar metallicity is $\sim 8~M_\odot$ \citep{smarttiip,heger}.
Meanwhile, the maximum ZAMS mass of Type IIP SN progenitors from
theoretical calculations of solar-metallicity single star evolution is
$\sim 25~M_\odot$ \citep[][and the references therein]{heger} but the observations suggest that the
maximum ZAMS mass is $\sim 17~M_\odot$ \citep{smarttiip}.

This big discrepancy partly comes from uncertainties in theoretical models.
One of the big uncertainties in modeling stellar evolution is in mass loss.
The canonical single star modeling uses empirical
mass-loss rates but 
there are many mechanisms to induce mass loss which are not taken into
account in such empirical mass-loss rates.
For example, the effect of stellar rotations on mass-loss rates
are studied for a long time and it is found that
the maximum ZAMS mass of the Type IIP SNe can be reduced
by including stellar rotations
\citep[][and the references therein]{georgy}.
Mass loss due to the pulsations driven by partial ionization of hydrogen
in the envelope of RSGs \citep[e.g.,][]{yoonRSG,hegerRSG,Li1994}
is another candidate for the missing mass-loss mechanism.
This kind of mass loss due to pulsations occurs in a dynamical timescale
and it is usually not followed by stellar evolution codes.
\citet{yoonRSG} analyse the dynamical instabilities of
RSGs and show that RSGs whose ZAMS mass is above $\sim17~M_\odot$
suffer from this instability and their mass-loss rate can potentially be
as high as $\sim 10^{-2}~M_\odot~\mathrm{yr^{-1}}$.
Such existence of extensive mass loss in massive RSGs
may not allow massive RSGs to explode as Type IIP SNe
and could reduce the maximum ZAMS mass of Type IIP SN progenitors.
It is worth noting that the minimum ZAMS mass
to cause the pulsations is the same as the maximum ZAMS mass
of Type IIP SN progenitors indicated by observations ($\sim 17~M_\odot$).
Nuclear flashes which may occur during the evolution of massive stars
can also be a candidate for the driving force of such extensive mass
loss \citep{neflash,lucmassloss}.
If such extensive mass loss takes place commonly during the evolution
of massive stars,
theoretical predictions for the maximum ZAMS
mass of Type IIP SN progenitors can be as low as what
observations imply.
The existence of binary systems is also suggested to
be account for this problem \citep[e.g.,][]{loss}.

There are also observed potential Type IIP SN progenitors (RSGs)
which are losing their mass with very high mass-loss rates.
For example, a RSG VY Canis Majoris
is estimated to be losing its mass 
with the rate $1-2\times10^{-3}~M_\odot~\mathrm{yr^{-1}}$
from the direct observations of circumstellar medium (CSM) around it \citep{smithRSG}.
Observations of dusts around another RSG IRAS05280-6910 also indicate the
extensive mass loss ($\sim 10^{-3}~M_\odot~\mathrm{yr^{-1}}$) of the RSG
\citep{har}.
Some Type IIP SNe are reported to show the possible effect of
dense CSM in their LCs and spectra.
A representative example of this kind is
ultraviolet(UV)-bright Type IIP SN 2009kf \citep{09kf}.
SN 2009kf was bright in UV for $\sim 10$ days during its early epochs
as well as in optical. Later, the LC transformed to that of a Type IIP SN
and the spectra taken at later epochs are classified as Type IIP.
The fact that SN 2009kf was bright in UV and optical at the same time
makes it difficult to explain SN 2009kf without CSM interaction.
This is because usual Type IIP SNe cool adiabatically in their early epochs
and they become optically bright after UV brightness gets weaker.
\citet{utrobin09kf} show a LC model of SN 2009kf without CSM
interaction but the required explosion energy is found to be very high
($2.2\times10^{52}~\mathrm{erg}$).
Another Type IIP SN 2007od \citep{07od} had late phase spectra similar
to those of Type IIn SN 1998S, which are considered to result
from CSM interaction \citep{98S}.
Also, SN 2007od showed the possibility of the existence of light echos
in its LC and it may indicate the existence of CSM around
the progenitor \citep{07od}.

The existence of dense CSM around
SN progenitors at the time of their explosions can affect their LCs. 
If a SN explosion occur in dense CSM, the shock breakout signal from the SN
diffuse out in CSM. In addition,
the SN ejecta is decelerated by the dense CSM.
Thus, the kinetic energy of the ejecta is efficiently
converted to the thermal energy which
is eventually released as radiation energy.
This extra heat source makes the SN brighter than usual SNe.
So far, SNe powered by
the interactions of SN ejecta and dense CSM are mainly modeled
in the context of Type IIn SNe showing narrow emission lines which are a clear
evidence of the interaction of SN ejecta and CSM
\citep[e.g.,][]{94w,06gy,luciin,iin1,iin2}.
Here, given the possibilities that RSGs can have
enhanced mass-loss rates
and that dense CSM can exist around RSGs at the time
of the explosions, we investigate
the effect of dense CSM on LCs from SNe of RSGs with dense CSM.
Some works \citep[e.g.,][]{falk73,falk77}
 have already done with the similar conditions
 but, in this paper, we model them more systematically and with better treatments of physics.
To model the LCs powered by the shock interactions,
we have to follow radiation hydrodynamics
because the energy source of the radiation is a hydrodynamical
shock wave. Thus, we adopt
a multi-group radiation hydrodynamics code
\verb|STELLA| developed by \citet{sergei93,sergei98,sergei06},
which has been applied for modeling LCs
powered by shock interactions \citep[e.g.,][]{94w,06gy,sergei10}.

This paper is composed of five sections.
We first briefly summarize \verb|STELLA| and the pre-SN models
we adopt in Section \ref{sec2}.
Synthesized LCs with various CSM
conditions are shown in Section \ref{LCs}.
In Section \ref{discussion}, we compare our models with the LCs of SN 2009kf and
show that the LCs powered by the interaction of SN ejecta
and dense CSM is actually consistent with the LCs of SN 2009kf.
We also get some implications for the CSM around the progenitor
of SN 2009kf and what kind of mass loss should have happened
before the explosion of SN 2009kf. Conclusions are given in Section \ref{conclusions}.

\begin{table}
\centering
\begin{minipage}{140mm}
\caption{List of LC models without CSM}
\label{nocsmlist}
\begin{tabular}{cccc}
\hline
Name & Progenitor & Explosion Energy & Radiation Energy\footnote{
Radiation energy emitted in 50 days since the explosion.}\\
     &            & $10^{51}$ erg & $10^{49}$ erg \\
\hline
s13e3 & s13 & 3 & 3.0 \\
s15e1 & s15 & 1 & 1.3 \\
s15e3 & s15 & 3 & 3.2 \\
s15e5 & s15 & 5 & 4.9 \\
s18e3 & s18 & 3 & 3.7 \\
s20e3 & s20 & 3 & 4.2 \\
\hline
\end{tabular}
\end{minipage}
\end{table}

\section{STELLA Code and Pre-Supernova Models}\label{sec2}
\verb|STELLA| is a one-dimensional multi-group radiation
hydrodynamics code \citep{sergei93,sergei98,sergei06}.
\verb|STELLA| calculates the spectral energy distributions
(SEDs) at each time step and we can get multicolor LCs by
convolving filter functions to the SEDs.
All the calculations are
performed by adopting 100 frequency bins from $\mathrm{1~\AA}$ to
$\mathrm{5\times10^{4} \AA}$.
\verb|STELLA| implicitly treats time-dependent equations
of the angular moments of intensity averaged over a frequency bin.
Local thermodynamic equilibrium is assumed to determine the
ionization levels of materials.
For the details of \verb|STELLA|, see \citet{sergei06} and the
references therein.
Comparisons of \verb|STELLA| with other numerical codes are provided in,
e.g., \citet{sergei98,sergei03,woosley07} and
analytical models are also compared to the numerical results of \verb|STELLA|
\citep[e.g.,][]{rabinak}.
In this paper, explosions are treated as a thermal bomb
by injecting thermal energy just below the mass cut $1.4~M_\odot$.
We do not follow explosive nucleosynthesis because
we focus on the early epochs when the effect of explosive
nucleosynthesis on SN LCs is negligible.

We construct pre-SN models by attaching CSM to the progenitor models
calculated by \citet{progenitor}.
As the progenitor models of \citet{progenitor} do not
take into account of the extensive mass loss we are interested in,
we artificially attach CSM to the outer most layer of the progenitor
models.
The composition of CSM is solar metallicity and is the same as the surface
of the RSG models which we adopt.
Compared to the effect of the CSM parameters,
the LCs are less affected by the RSG models adopted
inside (Section \ref{progenitor}).
This justifies our simple way to construct the pre-SN
models.
Among the pre-SN models shown in \citet{progenitor},
we use the solar-metallicity single star models of
RSGs; s13, s15, s18, and s20.
The ZAMS masses of the models s13, s15, s18, and s20 are
13 \Msun, 15 \Msun, 18 \Msun, and 20 \Msun, respectively.
If they are exploded without CSM, their LCs
show a long plateau phase and thus they are
Type IIP SN progenitors (Section \ref{nocsm}).
Although we do not follow explosive nucleosynthesis,
\Ni~is in the core of the pre-SN models due to the
nuclear statistical equilibrium established in the core.
However, at the early epochs we are interested in
(first $\sim50$ days since the explosions),
the photons originated from \Ni~decay do not leak from
the ejecta so much and contribute little to LCs.
According to Figure 2 of \citet{kaseniip},
the effect of \Ni~on Type IIP SN LCs
appear after $\sim50$ days since the explosions
and thus our LC models are not applicable from around that epoch.
Density structures of the pre-SN models
are shown accordingly in the following sections.

\begin{figure}
\begin{center}
 \includegraphics[width=\columnwidth]{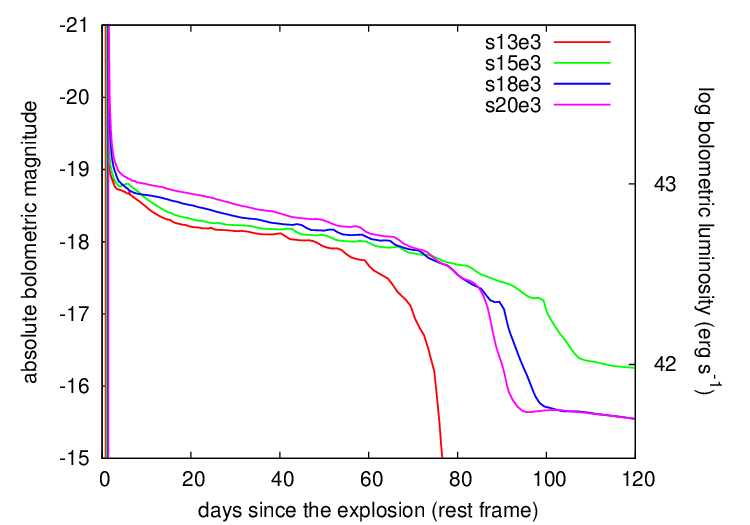}
  \caption{
Bolometric LCs of the SNe from the progenitors without CSM.
}
\label{sXXe3bol}
\end{center}
\end{figure}

\section{Light Curves}\label{LCs}
In this section, we present the LCs calculated by \verb|STELLA|.
First, we show LCs of the explosions of the progenitors without CSM
for references (Section \ref{nocsm}) and then
LCs of the explosions with CSM (Section \ref{wcsm}).

\begin{figure*}
\begin{center}
 \includegraphics[width=\columnwidth]{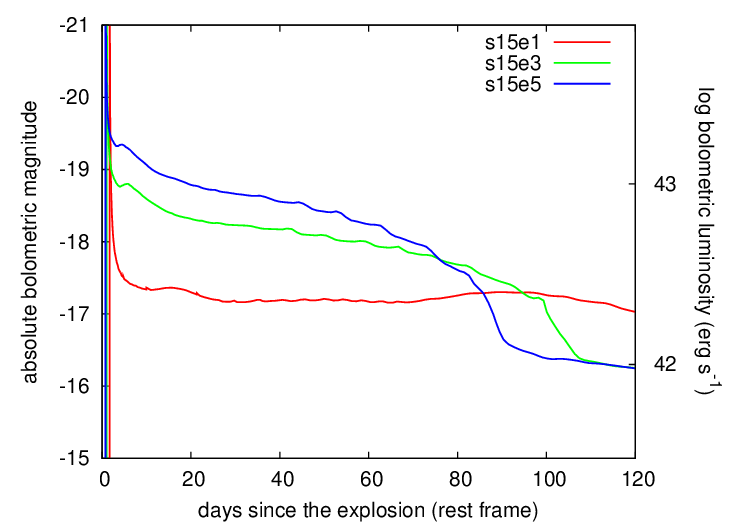}
 \includegraphics[width=\columnwidth]{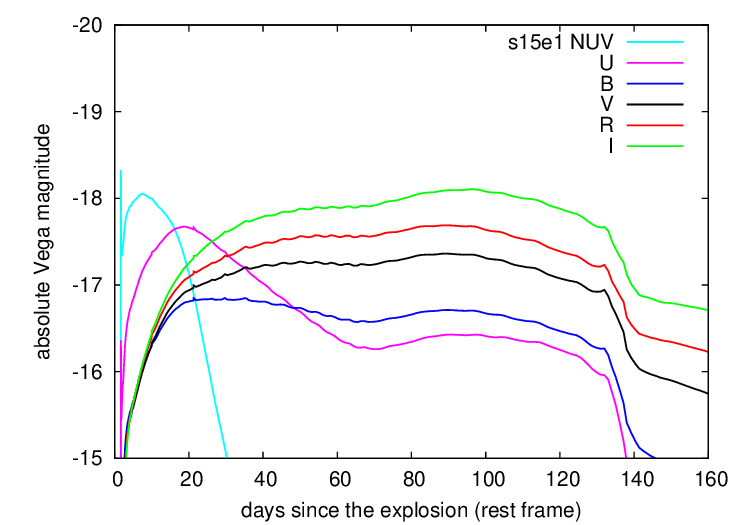}\\
 \includegraphics[width=\columnwidth]{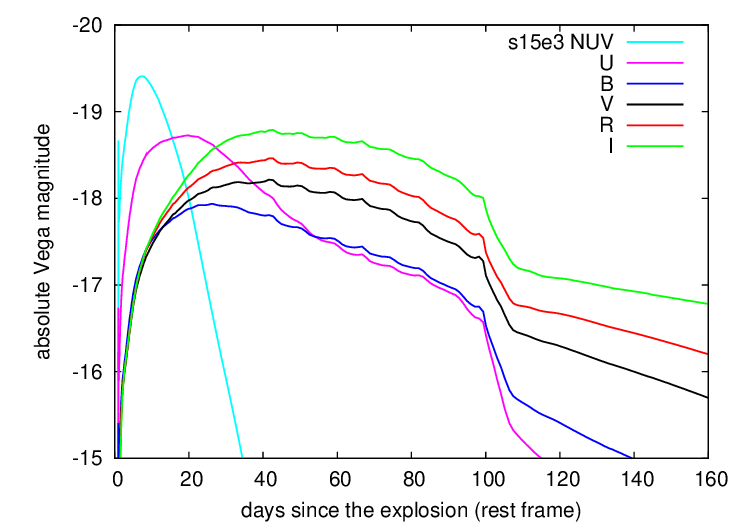}
 \includegraphics[width=\columnwidth]{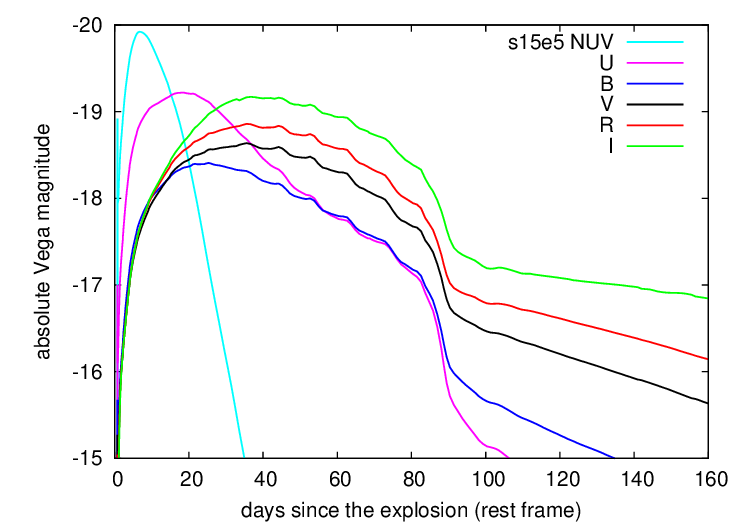}
 \caption{
Bolometric (top left) and multicolor (the others) LCs of the SNe from
the progenitor s15 without CSM.
Models with several explosion energies are shown.
}
\label{s15eX}
\end{center}
\end{figure*}

\subsection{Explosions without CSM}\label{nocsm}
In this section, we present the LCs
for the progenitors without CSM.
The aim of these calculations
is to provide references to be compared with
LCs with CSM.
Note that, as mentioned in Section \ref{sec2},
the LCs shown here are not applicable after $\sim 50$ days
since the explosions because we do not follow explosive nucleosynthesis.
Table \ref{nocsmlist} is the list of the LCs shown in this section.
Previous studies also calculated the multicolor LCs of Type IIP SNe from the
same progenitor models with different numerical codes
\citep{kaseniip,luciip,luciip2} and
from different progenitor models with the same numerical code
\citep{baklanov2005,tominagaiip,tominagashbr}.

In Figure \ref{sXXe3bol}, the bolometric LCs
with the same explosion energy ($3\times10^{51}$ erg) are presented.
When a shock emerges from a stellar surface,
a bolometric LC is suddenly brightened
due to the shock breakout. Then, the ejecta cools adiabatically.
When the outer layer of the ejecta reaches the recombination temperature
of hydrogen, the LC becomes flat until the photosphere reaches
the bottom of the hydrogen layer (plateau phase).
After the plateau phase, the LC follows the decay line of $\mathrm{^{56}Co}$
which existed as \Ni~in the core of the pre-SN model
even though we do not calculate the explosive nucleosynthesis (Section \ref{sec2}).
Note that our LC models are not applicable after the late epochs of the plateau phase.

The bolometric LCs with the same progenitor (s15) but different explosion energies
are shown in the top left panel of Figure \ref{s15eX}.
In the other panel of Figure \ref{s15eX}, we also show the multicolor LCs of each model.
Optical LCs are obtained by applying the Bessell
$UBVRI$ filters \citep{bessell} and UV LCs are derived by using the near-UV (NUV) imaging filter
of the {\it Galaxy Evolution Explorer} ({\it GALEX}) satellite whose
central wavelength is around $2300~\mathrm{\AA}$ \citep{galex}.
After the first brightening due to shock breakout,
NUV first become bright because of the adiabatic cooling of the ejecta.
Then, LCs become bright in optical as NUV become fainter.
This is an important feature of usual Type IIP SNe. They are not bright
in UV and optical at the same time.
The behavior of the multicolor LCs of the other progenitors is qualitatively
the same as the multicolor LCs shown in Figure \ref{s15eX}.

\subsection{Explosions with CSM}\label{wcsm}
\begin{figure}
\begin{center}
 \includegraphics[width=\columnwidth]{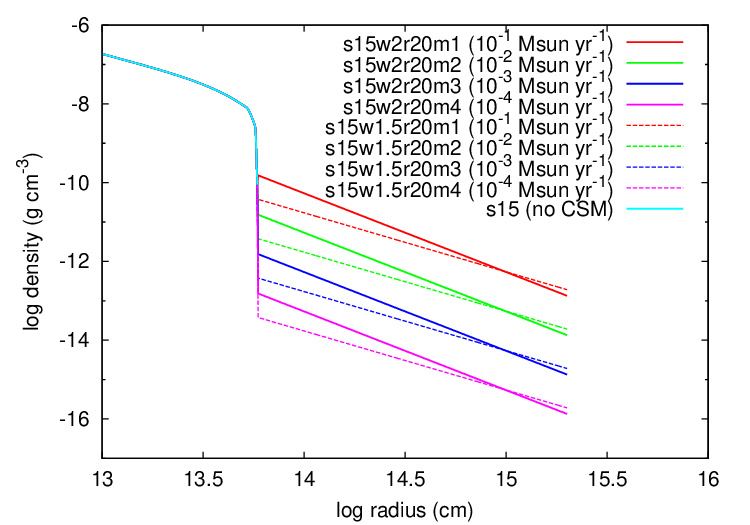}
 \caption{
Density structures of the pre-SN models
with different mass-loss rates and density slopes.
The density structures shown with solid lines have
CSM density slope of $\rho\propto r^{-2}$
and those shown with dashed lines have
CSM density slope of $\rho\propto r^{-1.5}$.
The pre-SN models are constructed by attaching CSM to
the progenitor model s15.
}
\label{masslossdensity}
\end{center}
\end{figure}

In this section, we investigate the effect of CSM around RSGs on SN LCs.
After the explosion of the progenitor, a shock wave propagates inside the progenitor.
At these epochs, CSM is not ionized and optically thin.
Then, after the shock wave has gone through the surface of the progenitor,
a precursor wave appears in the CSM and CSM is ionized.
The precursor wave propagates very fast and soon reaches to the surface of CSM.
The optical depth ($\tau_{\mathrm{CSM}}$) of CSM after the shock wave reached the
surface of the progenitor is expressed as
\begin{eqnarray}
\tau_{\mathrm{CSM}}=\int^{R}_{R_0}\kappa(r)\rho(r) dr
\simeq\int^{R_i}_{R_0}\kappa(r)\rho(R_0)\left(\frac{r}{R_0}\right)^{-\alpha}dr, \label{eq1}
\end{eqnarray}
where $\kappa$ is opacity, $\rho$ is density, $R$ is the CSM radius,
$R_0$ is the radius of
the inner edge of CSM where CSM is connected to the progenitor inside, i.e.,
the radius of the progenitor inside, and $R_i$ is the ionization front in CSM.
$\tau_\mathrm{CSM}$ depends on the thickness of the ionized layer because
$\kappa$ above the ionization front is very low $(\kappa\sim10^{-4}~\mathrm{cm^{2}~g^{-1}})$.
In order to estimate a condition in which CSM can affect LCs,
we assume that CSM below $R_i$ is fully ionized and
the Thomson scattering is the predominant source of opacity
($\kappa(r)=0.33~\mathrm{cm^2~g^{-1}}$ with solar metallicity).
With $R_0\simeq 5\times 10^{13}~\mathrm{cm}$ (Figure \ref{masslossdensity};
Section \ref{progenitor}),
$R_0\ll R_i$, and $\alpha\hspace{0.3em}\raisebox{0.4ex}{$>$}\hspace{-0.75em}\raisebox{-.7ex}{$\sim$}\hspace{0.3em} 1$, the condition to be $\tau_\mathrm{CSM}>1$ is
\begin{eqnarray}\label{opacity}
\rho (R_0) 
\hspace{0.3em}\raisebox{0.4ex}{$>$}\hspace{-0.75em}\raisebox{-.7ex}{$\sim$}\hspace{0.3em} 
6\times 10^{-14} \left( \alpha-1 \right)~\mathrm{g~cm^{-3}}.
\end{eqnarray}
Thus, when the density of CSM at the radius where CSM is connected
to the progenitor model inside is more than 
$\sim10^{-13}~\mathrm{g~cm^{-3}}$, CSM becomes optically thick and
LCs are expected to be affected by CSM.
CSM with mass-loss rates higher than $\sim10^{-4}~M_\odot~\mathrm{yr^{-1}}$
satisfies this condition, assuming that the CSM velocity is $10^{6}~\mathrm{cm~s^{-1}}$
(Figure \ref{masslossdensity}).

In the following sections,
we also investigate the dependence of LCs on several physical parameters of CSM and progenitors.
The parameters of CSM adopted are
mass-loss rates [$10^{-1}-10^{-4}M_\odot~\mathrm{yr^{-1}}$],
radii of the outer edge of CSM (CSM radii)
[$5\times10^{14}-3\times10^{15}\mathrm{cm}$],
and density slopes ($\rho\propto r^{-\alpha}$)  [$\alpha=2, 1.5$].
Several progenitor models inside are also adopted [s13, s15, s18, s20].
The density structures of the pre-SN models with CSM
are shown accordingly in the following sections.
The model parameters and results are summarized in Tables \ref{wcsmlist} and \ref{wcsmprop}.

\subsubsection{Effect of CSM}\label{effect}

\begin{figure}
\begin{center}
 \includegraphics[width=\columnwidth]{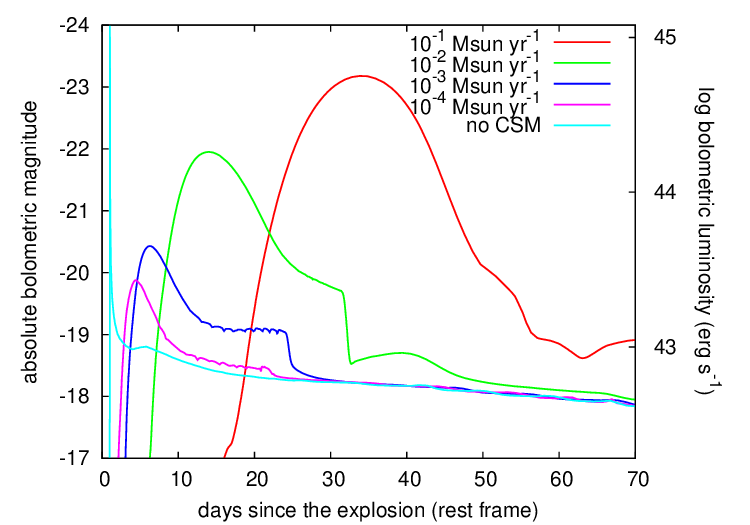}
 \caption{
Bolometric LCs of the models with different mass-loss rates.
The models shown are s15w2r20m1e3 ($10^{-1}~M_\odot~\mathrm{yr^{-1}}$),
s15w2r20m2e3 ($10^{-2}~M_\odot~\mathrm{yr^{-1}}$),
s15w2r20m3e3 ($10^{-3}~M_\odot~\mathrm{yr^{-1}}$), and
s15w2r20m4e3 ($10^{-4}~M_\odot~\mathrm{yr^{-1}}$).
The bolometric LC of the model s15e3 (no CSM) is also shown for comparison.
}
\label{masslossbol}
\end{center}
\end{figure}

Typical bolometric LCs which are affected by CSM are shown in Figure \ref{masslossbol}
with the bolometric LC from the model without CSM (s15e3).
We focus on the model s15w2r20m3e3 to describe the effect of CSM in this section.
CSM mainly affects the LCs at the early epochs, roughly until the sudden drop in the LCs
which can be seen at around 25 days in the model s15w2r20m3e3
($10^{-3}~M_\odot~\mathrm{yr^{-1}}$).
At first, LCs have round shapes (until $\sim 15$ days in s15w2r20m3e3).
The round phase is followed by a flat LC which lasts until the sudden drop
($\sim15-25$ days in s15w2r20m3e3).
LCs are mainly powered by the interaction of SN ejecta and CSM at these epochs.
We call these epochs as an interaction-powered phase (IPP).

\begin{figure}
\begin{center}
 \includegraphics[width=\columnwidth]{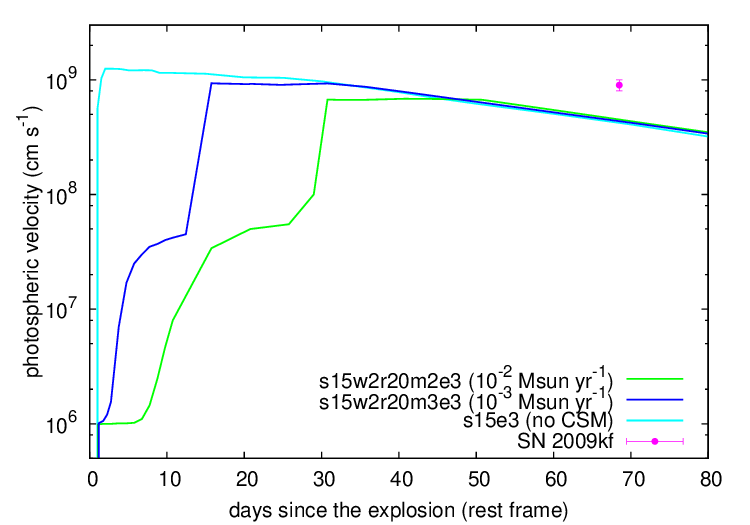}
 \caption{
Evolutions of photospheric velocity of the models s15w2r20m2e3, s15w2r20m3e3, and s15e3.
The photosphere is defined as the location where the Rosseland optical depth
becomes 2/3. Observed H$\alpha$ line velocity of SN 2009kf is also plotted in the figure.
The explosion date of SN 2009kf is set as the same as in Figure \ref{09kfcomp}.
}
\label{photov}
\end{center}
\end{figure}

\begin{figure*}
\begin{center}
 \includegraphics[width=\columnwidth]{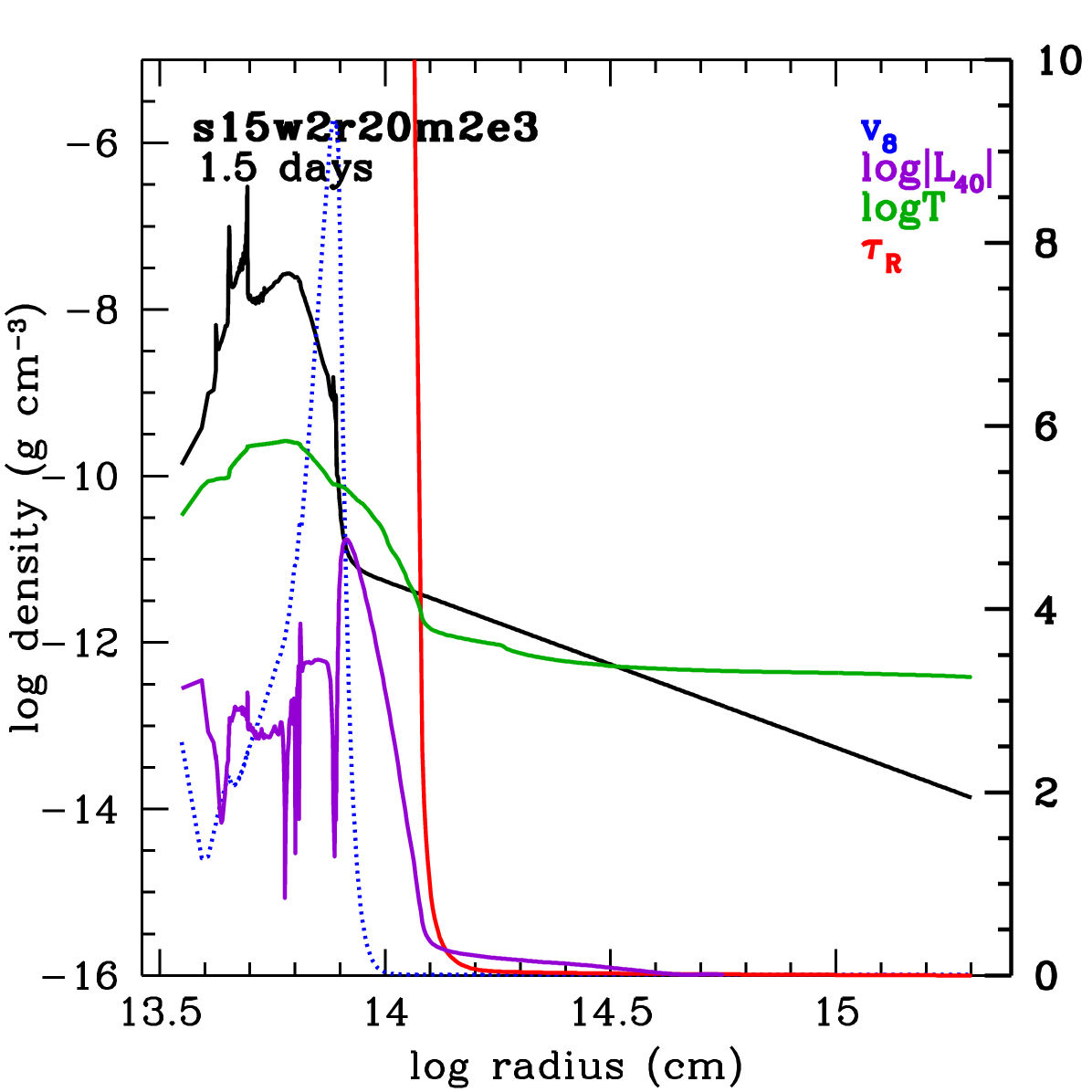}
 \includegraphics[width=\columnwidth]{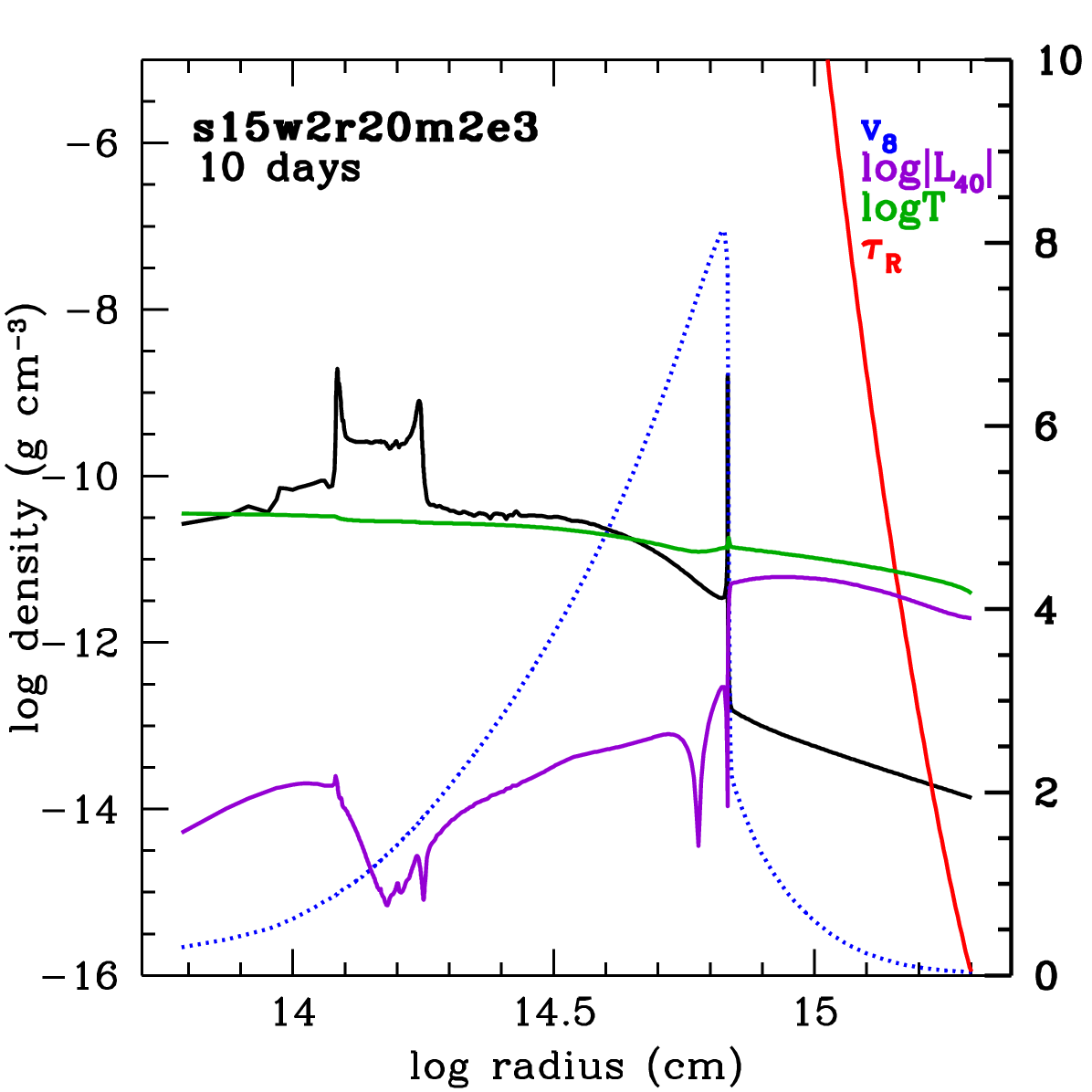}\\
 \includegraphics[width=\columnwidth]{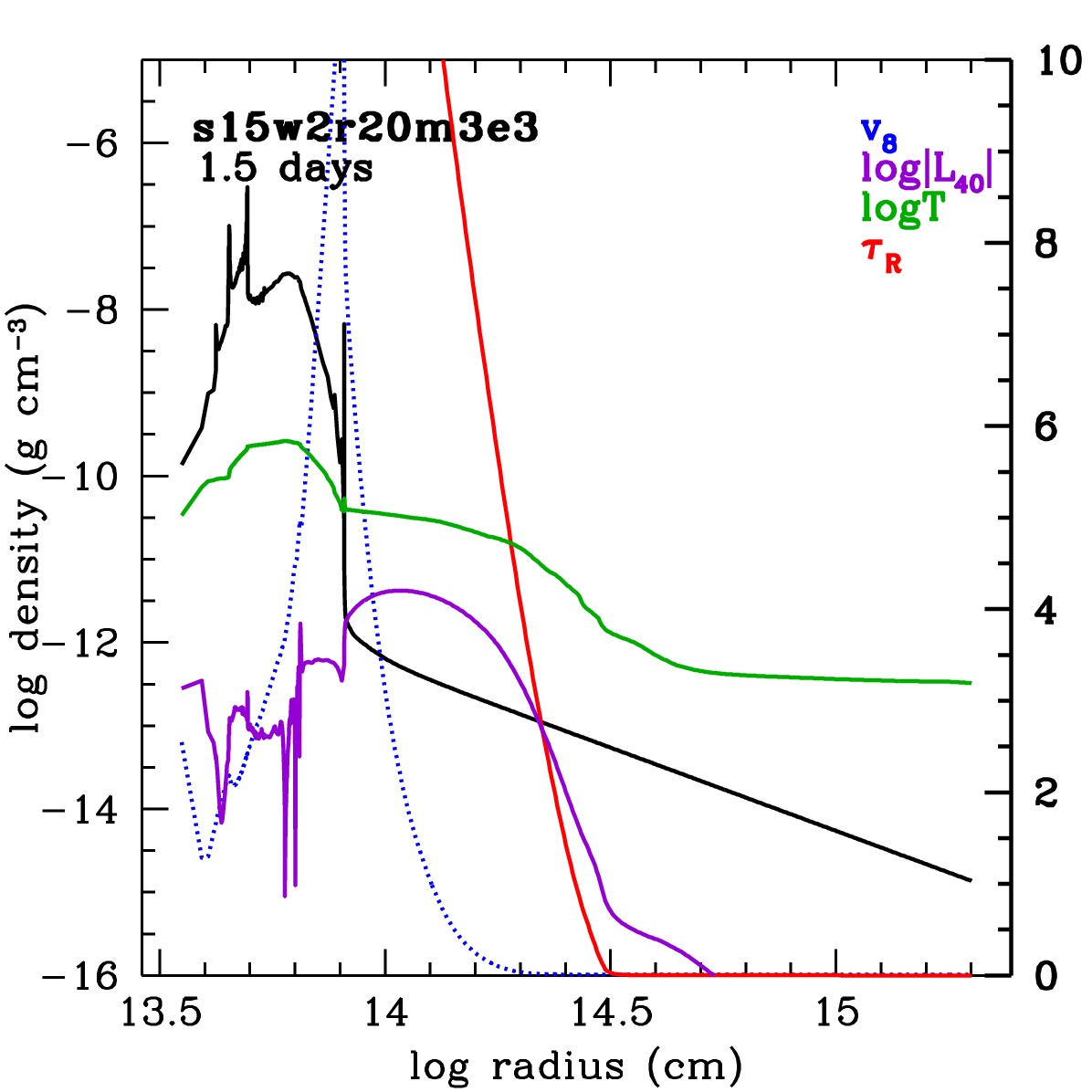}
 \includegraphics[width=\columnwidth]{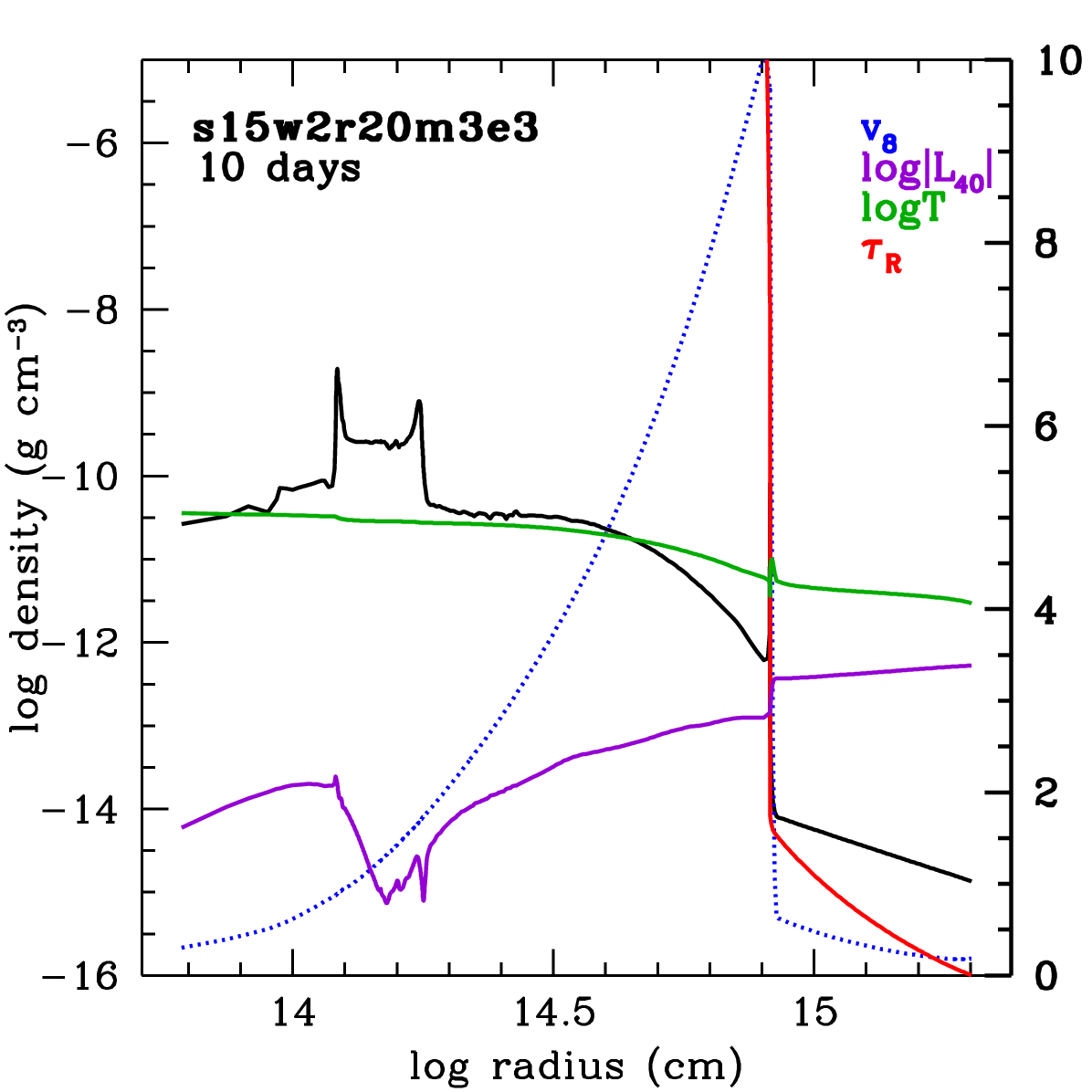}
 \caption{
Physical structures of the models s15w2r20m2e3 and s15w2r20m3e3
at two epochs. The epochs are the days since the explosion.
Black lines show the density structure (left $y$-axis).
Blue lines are the velocity scaled by $10^{8}~\mathrm{cm~s^{-1}}$ (right
$y$-axis), purple lines are the logarithm of the
absolute value of luminosity scaled by $10^{40}~\mathrm{erg~s^{-1}}$
 (right $y$-axis),
green lines are the logarithm of the temperature in Kelvin (right
 $y$-axis), and
red lines are Rosseland optical depth measured from the outside
(right $y$-axis).
}
\label{masslossevol}
\end{center}
\end{figure*}

The IPP appears in the LCs of the SNe with dense CSM.
When the temperature and CSM density are high enough, CSM becomes optically thick and the photosphere locates in CSM. 
Photosphere during the IPP is in CSM. This can clearly seen
in Figure \ref{photov}. Looking into the model s15w2r20m3e3, the photospheric velocity is
at first $10^{6}~\mathrm{cm~s^{-1}}$, which is the initial CSM velocity.
This is a characteristic feature of the explosions within dense CSM.
Then, the photospheric velocity increases due to the acceleration of CSM by the precursor wave.
At around $\sim15$ days in s15w2r20m3e3, the photosphere reaches the shell between
SN ejecta and CSM (Figure \ref{masslossevol}).
Then, the photosphere goes into SN ejecta at $\sim25$ days.

\begin{table*}
\centering
\begin{minipage}{180mm}
\caption{List of LC models with CSM}
\label{wcsmlist}
\begin{tabular}{cccccccc}
\hline
Name  & Progenitor & Explosion Energy &
Mass-Loss Rate\footnote{Derived by assuming a constant CSM velocity
$10^{6}~\mathrm{cm~s^{-1}}$.}
& Radius & Density Slope & CSM Mass & Radiation Energy\footnote{
Radiation energy emitted in 50 days since the explosion.}
\\
&& $10^{51}$ erg &  $M_\odot~\mathrm{yr^{-1}}$ &
$10^{15}$ cm& $\alpha~(\rho\propto r^{-\alpha})$ & $M_\odot$
& $10^{49}$ erg
\\
\hline
s13w2r20m2e3 & s13 & 3 & $10^{-2}$ & 2 & 2 & 0.65 & 20 \\
s15w2r05m2e3 & s15 & 3 & $10^{-2}$ & 0.5 & 2 & 0.15 & 9.8 \\
s15w2r05m3e3 & s15 & 3 & $10^{-3}$ & 0.5 & 2 & 0.015 & 3.8 \\
s15w2r10m2e3 & s15 & 3 & $10^{-2}$ & 1 & 2 & 0.31 & 13 \\
s15w2r10m3e3 & s15 & 3 & $10^{-3}$ & 1 & 2 & 0.031 & 4.2 \\
s15w2r20m2e1 & s15 & 1 & $10^{-2}$ & 2 & 2 & 0.65 & 5.6\\
s15w2r20m1e3 & s15 & 3 & $10^{-1}$ & 2 & 2 & 6.5 & 75 \\
s15w1.5r20m1e3 & s15 & 3 & $-$ & 2 & 1.5 & 6.5 & 80 \\
s15w2r20m2e3 & s15 & 3 & $10^{-2}$ & 2 & 2 & 0.65 & 19 \\
s15w1.5r20m2e3 & s15 & 3 & $-$ & 2 & 1.5 & 0.65 & 15 \\
s15w2r20m2e5 & s15 & 5 & $10^{-2}$ & 2 & 2 & 0.65 & 32 \\
s15w2r20m2e7 & s15 & 7 & $10^{-2}$ & 2 & 2 & 0.65 & 45 \\
s15w2r20m3e3 & s15 & 3 & $10^{-3}$ & 2 & 2 & 0.065 & 5.0 \\
s15w1.5r20m3e3 & s15 & 3 & $-$ & 2 & 1.5 & 0.065 & 4.8 \\
s15w2r20m4e3 & s15 & 3 & $10^{-4}$ & 2 & 2 & 0.0065 & 3.4 \\
s15w1.5r20m4e3 & s15 & 3 & $-$ & 2 & 1.5 & 0.0065 & 3.4 \\
s15w2r30m2e3 & s15 & 3 & $10^{-2}$ & 3 & 2 & 0.98 & 23 \\
s15w2r30m3e3 & s15 & 3 & $10^{-3}$ & 3 & 2 & 0.098 & 5.6 \\
s18w2r20m2e3 & s18 & 3 & $10^{-2}$ & 2 & 2 & 0.65 & 17 \\
s20w2r20m2e3 & s20 & 3 & $10^{-2}$ & 2 & 2 & 0.65 & 17 \\
\hline
\end{tabular}
\end{minipage}
\end{table*}

\begin{table*}
\centering
\begin{minipage}{140mm}
\caption{Properties of SNe from s15 at some selected epochs}
\label{wcsmprop}
\begin{tabular}{ccccccccccccc}
\hline
  & \multicolumn{4}{c}{$L_\mathrm{bol}$\footnote{Bolometric luminosity.} ($10^{43}~\mathrm{erg~s^{-1}}$)} &
\multicolumn{4}{c}{$v_\mathrm{ph}$\footnote{Photospheric velocity. Photosphere is where the Rosseland optical depth is 2/3.} ($\mathrm{km~s^{-1}}$)} &
\multicolumn{4}{c}{$T_\mathrm{BB}$\footnote{Color temperature at photosphere.
Photosphere is where the Rosseland optical depth is 2/3.} ($10^{3}~\mathrm{K}$)}\\
Epoch\footnote{Epoch since the explosion.} (days) &10 & 20  &  30  & 40 &
10 & 20  &  30  & 40  &
10 & 20  &  30  & 40 
\\
\hline
s15e1 &0.27&0.26&0.23&0.23&6900&6500&5900&5000&15&9.8&7.9&7.0\\
s15e3 &0.84&0.65&0.60&0.57&12000&11000&9500&7800&16&9.8&8.0&7.1\\ 
s15e5 &1.3&1.0&0.90&0.81&15000&13000&12000&9200&15&9.9&8.0&7.1\\
s15e7 &1.7&1.4&1.2&1.0&17000&16000&13000&10000&16&9.9&8.1&7.1\\
s15w2r05m2e3 &1.1&0.63&0.73&0.57&8100&8300&8300&7800&15&9.0&7.7&7.2 \\
s15w2r05m3e3 &0.82&0.69&0.61&0.56&10000&10000&9400&7800&13&9.6&8.0&7.2 \\
s15w2r10m2e3 &22&0.78&0.73&0.64&950&7800&7500&7600&42&9.1&7.9&7.2 \\
s15w2r10m3e3 &1.5&0.72&0.61&0.56&9800&9800&9400&7800&14&9.1&8.0&7.2 \\
s15w2r20m2e1 &0.81&3.2&1.2&0.67&14&86&130&150&15&21&14&10 \\
s15w2r20m1e3 &0.049&1.9&47&36&10&15&410&1200&6.0&13&37&32 \\
s15w1.5r20m1e3 &0.064&1.5&45&46&10&12&400&1400&6.0&12&36&34 \\
s15w2r20m2e3 &9.1&8.4&2.5&0.92&65&500&7200&6800&29&23&11&7.2 \\
s15w1.5r20m2e3 &5.0&7.9&3.0&0.64&43&290&6300&5600&23&25&16&7.5\\ 
s15w2r20m2e5 &29&7.6&1.2&1.1&670&1000&8900&9000&39&20&7.9&7.3 \\
s15w2r20m2e7 &58&8.1&1.7&1.3&570&11000&11000&10000&42&16&8.2&7.4 \\
s15w2r20m3e3 &2.1&1.3&0.62&0.57&420&9400&9200&7800&18&9.5&8.0&7.2 \\
s15w1.5r20m3e3 &2.1&1.5&0.62&0.36&270&9200&9200&7800&18&9.8&8.1&7.2\\ 
s15w2r20m4e3 &0.96&0.74&0.61&0.57&11000&10000&9400&7800&16&10&8.1&7.1 \\
s15w1.5r20m4e3 &0.90&0.77&0.61&0.56&11000&10000&9500&7800&17&11&8.1&7.2\\ 
s15w2r30m2e3 &2.8&15&4.4&2.5&21&210&300&330&20&26&16&10 \\
s15w2r30m3e3 &3.5&1.3&1.1&0.61&30&9400&9100&8000&21&9.8&8.7&7.3 \\
\hline
\end{tabular}
\end{minipage}
\end{table*}

All the LCs affected by CSM have round shapes at first (Figure \ref{masslossbol}).
We briefly discuss why the round phase appears.
As our models have $\tau_{\mathrm{CSM}}>1$, photons cannot escape freely from
CSM.
In addition, our models satisfy the following condition when the shock wave
is propagating in CSM:
\begin{equation}
\tau_s=\int_{R_s}^R\kappa(r)\rho(r)dr<\frac{c}{v_s},\label{diffco}
\end{equation}
where $R_s$ is the radius of the shock wave and $v_s$ is the shock velocity.
Therefore, photons can diffuse out from the shock wave and
a precursor wave propagates ahead of the shock wave
(see the left panels of Figure \ref{masslossevol}).
As the shock velocity is typically $\simeq 10^{9}~\mathrm{cm~s^{-1}}$
when the shock wave reach $R_0$, $c/v_s$ is typically $\sim 10$ at that time.
Thus, photons in the models with
\begin{equation}
\rho(R_0)
\hspace{0.3em}\raisebox{0.4ex}{$<$}\hspace{-0.75em}\raisebox{-.7ex}{$\sim$}\hspace{0.3em} 
6\times10^{-13}~\mathrm{g~cm^{-3}}
\end{equation}
start to leak photons from the shock wave just after the shock wave reach $R_0$.
The other models satisfy Equation \ref{diffco} when the shock wave propagates
in CSM because the deceleration of the shock wave makes $c/v_s$ higher
and the propagation of the shock wave reduces $\tau_s$.
Once Equation \ref{diffco} is satisfied, photons start to leak out from the shock wave and
this phenomenon is usually observed as shock breakout.
However, since there is dense CSM at the time of shock breakout, photons diffuse out in CSM
and shock breakout signal become longer compared with the explosions without CSM
\citep[see also][]{falk73,falk77}.

In addition, SN ejecta is decelerated by
the dense CSM. Dense 
CSM is massive and it has the density structure $\rho\propto r^{-2}$ if it is from steady mass loss.
Thus, the shock wave between the SN ejecta and CSM is decelerated and kinetic energy of
the SN ejecta is converted to thermal energy which is released as radiation energy.
As a result, SNe with dense CSM emit more photons and become
brighter than those without dense CSM.
For further discussion, see, e.g., \citet{che11}.
Comparing radiation energy emitted during early epochs
(Tables \ref{nocsmlist} and \ref{wcsmlist}), it is clear that
the effect of the SN ejecta deceleration is dominant radiation source during the IPP.
In other words, the round phase is not just due to the elongation of
the shock breakout signal seen in the models without CSM (e.g. s15e3).

\begin{figure*}
\begin{center}
 \includegraphics[width=\columnwidth]{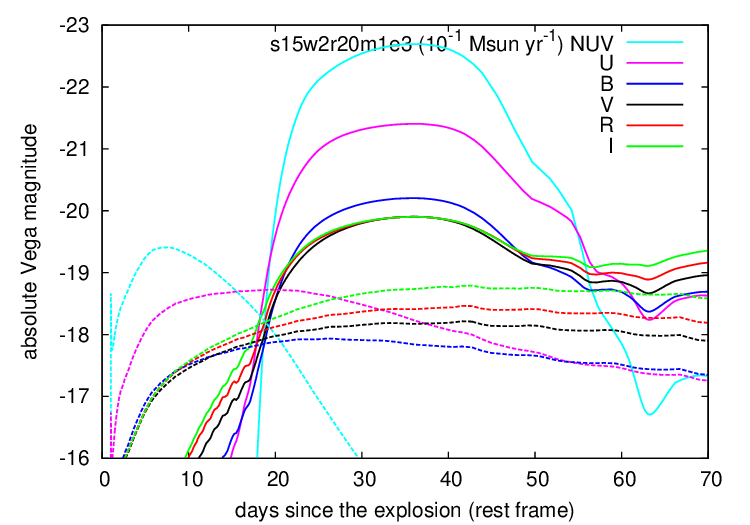}
 \includegraphics[width=\columnwidth]{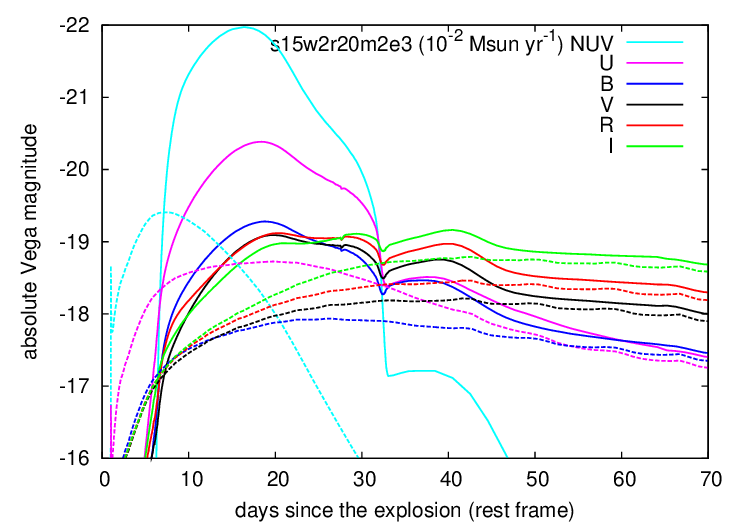}
 \includegraphics[width=\columnwidth]{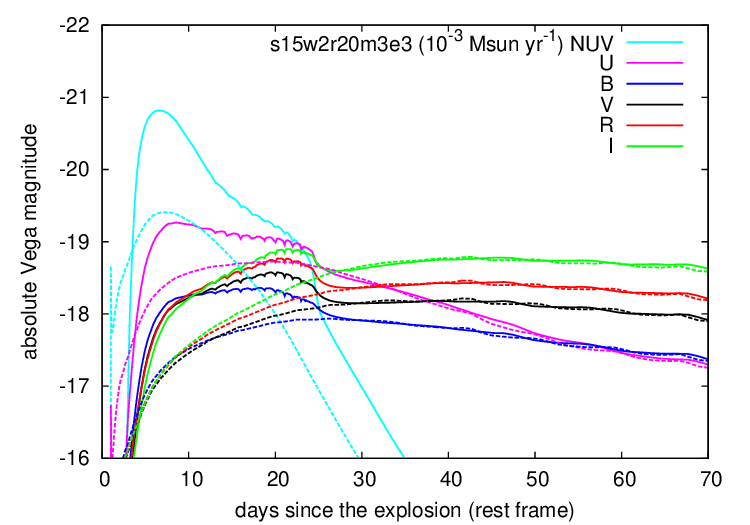}
 \includegraphics[width=\columnwidth]{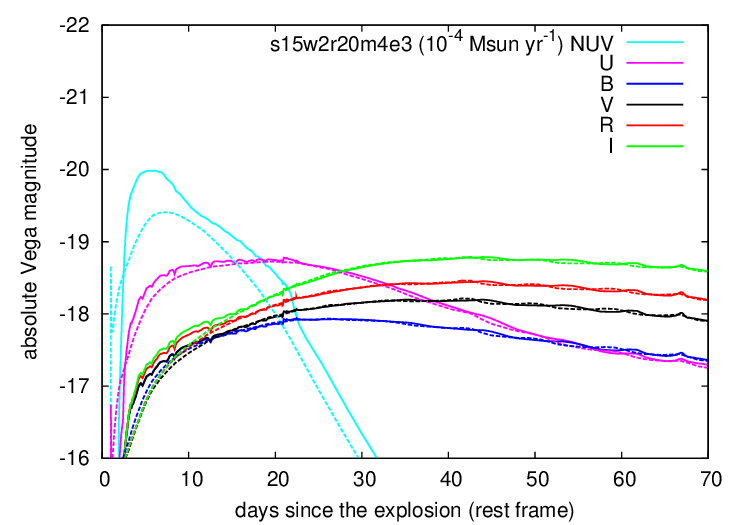}\\
 \caption{
Multicolor LCs of the models with different mass-loss rates (solid lines).
The dashed lines are the multicolor LCs of the model s15e3 for
 comparison. LCs of the same color are from the same filter.
 Note that top left panel has different $y$-axis range from the other panels.
}
\label{masslosscol}
\end{center}
\end{figure*}

The difference in the rising times and durations of the round LCs during the IPP
comes from the difference in diffusion timescales of CSM (Figure \ref{masslossbol}).
The models with higher mass-loss rates have denser CSM and thus longer diffusion timescales.
This difference due to diffusion timescales can be clearly seen in Figure \ref{masslossevol}.
The upper panels and the lower panels represent
the same epoch of the explosion with different CSM densities, i.e., different CSM diffusion timescales.
As is indicated by the temperature waves
in CSM which are pushed by the photon diffusion,
photons in CSM with shorter diffusion timescales diffuse out more quickly into
CSM.
Therefore, the rising times and the durations of the round phase in the IPP are shorter for
the models with the smaller CSM diffusion timescales.
Those differences are also discussed in \citet{falk73,falk77}.

The round LC in the IPP is followed by the flat LC which lasts until the sudden drop of the LC
(between around 15 days and 25 days in s15w2r20m3e3).
During this flat phase, the photosphere locates at the dense shell
between SN ejecta and CSM.
The photospheric velocity does not change so much during the flat phase (Figure \ref{photov}).
After CSM above the dense shell
have become optically thin, the photosphere remains at the dense shell
until the temperature and density of the shell becomes low enough to be optically thin.

One of the clear characteristics of the LCs with CSM is the
sudden drop in the LCs.
The time of this sudden drop corresponds to the time when the dense shell
becomes optically thin and the photosphere proceeds inward to SN ejecta.
This can also be seen in the photospheric velocity evolution (Figure \ref{photov}).
The brightness can drop as low as the LC without CSM
because now the photons are emitted
from SN ejecta and the physical conditions are the same as
those of SNe without dense CSM.
However,  the brightness is still slightly more luminous
than the LC without CSM for several days after the
sudden drop.
This could be because of the extra-heating due to shock and/or
deceleration of SN ejecta by CSM which makes the adiabatic cooling of
SN ejecta less efficient.

There are many differences between our models (RSG + CSM) and RSGs with extended
envelopes. First of all, it is difficult to have RSGs extended to $\sim 10^{15}$ cm
\citep[e.g.,][]{progenitor}.
What is more, density and temperature in RSGs are much higher in RSG envelopes than in CSM.
This does not allow the shock wave in RSGs to satisfy Equation \ref{diffco} until
the shock wave reaches the surface of RSGs. In other words, the shock wave does not break out
until it reaches near the surface.
On the contrary, in our models, Equation \ref{diffco} is satisfied
inside CSM and the precursor wave propagates ahead of the shock wave.
In addition, the kinetic energy of the SN ejecta efficiently converted to radiation energy because of 
the deceleration of the SN ejecta by the dense CSM.
This precursor wave due to the shock breakout in CSM and an additional heating source
causes the IPP phase.

\begin{figure}
\begin{center}
 \includegraphics[width=\columnwidth]{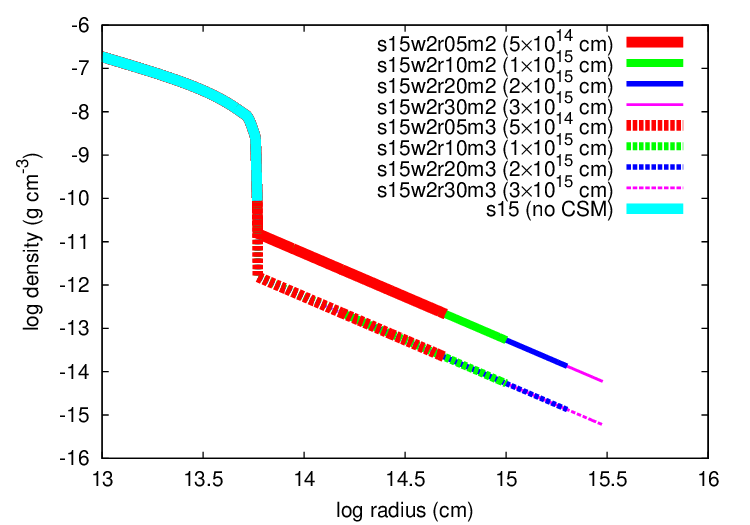}
 \caption{
Density structures of the pre-SN models with
different radii.
Models with solid lines have the mass-loss rate $10^{-2}~M_\odot~\mathrm{yr^{-1}}$ and
those with dashed lines have $10^{-3}~M_\odot~\mathrm{yr^{-1}}$.
The pre-SN models are constructed by attaching CSM
to the progenitor model s15.
}
\label{radiidensity}
\end{center}
\end{figure}

\subsubsection{Dependence on Mass-Loss Rate}\label{massloss}
In this section, we show the effect of mass-loss rates on LCs.
We adopt the mass-loss rates of $10^{-1}$, $10^{-2}$, $10^{-3}$, and $10^{-4}$
$M_\odot~\mathrm{yr^{-1}}$
(the corresponding model names are
s15w2r20m1, s15w2r20m2, s15w2r20m3, and s15w2r20m4, respectively).
For the case of $10^{-1}~M_\odot~\mathrm{yr^{-1}}$,
CSM mass is $6.5~M_\odot$ and the sum of the mass of RSG inside and CSM exceeds
the ZAMS mass of the progenitor. This is unrealistic but we show the results just
to see the effect of CSM.
Every CSM in the models is optically thick (Equation \ref{opacity}; Figure \ref{masslossdensity}).
These mass-loss rates are derived by assuming that the
CSM velocity is $10^{6}~\mathrm{cm~s^{-1}}$.
However, the escape velocity of the surface of s15 is 
$7.6\times10^{6}~\mathrm{cm~s^{-1}}$.
This means that the CSM velocity can be higher than
$10^{6}~\mathrm{cm~s^{-1}}$
at least at the late stages of the evolution of the progenitor
and thus the actual mass-loss rate for a given CSM
density could be higher than the values we show.
The flow from the progenitor may not be even steady.
However, we assume that CSM results from the steady flow from
the progenitor with the velocity $10^{6}~\mathrm{cm~s^{-1}}$
for simplicity because the CSM velocity has little effect on LCs.
To see the effect of the mass-loss rates on the LCs,
we fix the radius of CSM to $2\times10^{15}~\mathrm{cm}$
and the density slope to $\rho\propto r^{-2}$ in this section.
Also, the explosion energy and the progenitor of the SNe
are fixed to $3\times10^{51}$ erg and s15.
The density structures of the progenitors with CSM are
shown in Figure \ref{masslossdensity}.

Figure \ref{masslossbol} shows the bolometric LCs of the SNe
with different mass-loss rates.
Since the diffusion timescale of CSM becomes larger for denser CSM,
rising times and durations of round phases in LCs are longer for the models
with higher mass-loss rates.
The maximum luminosity of the LCs becomes larger
as CSM becomes denser.
This is because
the shock wave is more decelerated by CSM
due to the more massive CSM, i.e.,
more kinetic energy is converted
to the thermal energy and thus radiation energy.
The radiation energies of the models are summarized in Table \ref{wcsmlist}.

The multicolor LCs of the models in this section
are shown in Figure \ref{masslosscol}.
Each LC is plotted with the multicolor LCs of
no CSM model s15e3.
It is clear that the LCs during the IPP become bright
especially in short wavelengths because of high temperatures
at the photosphere.

\begin{figure}
\begin{center}
 \includegraphics[width=\columnwidth]{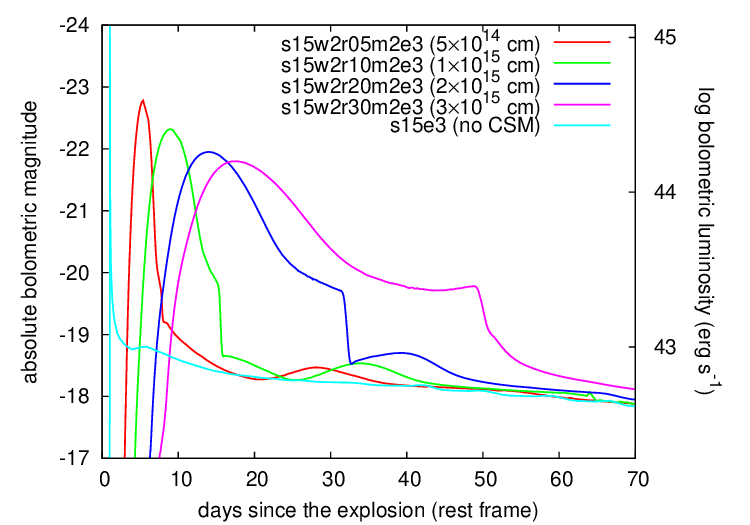}
  \includegraphics[width=\columnwidth]{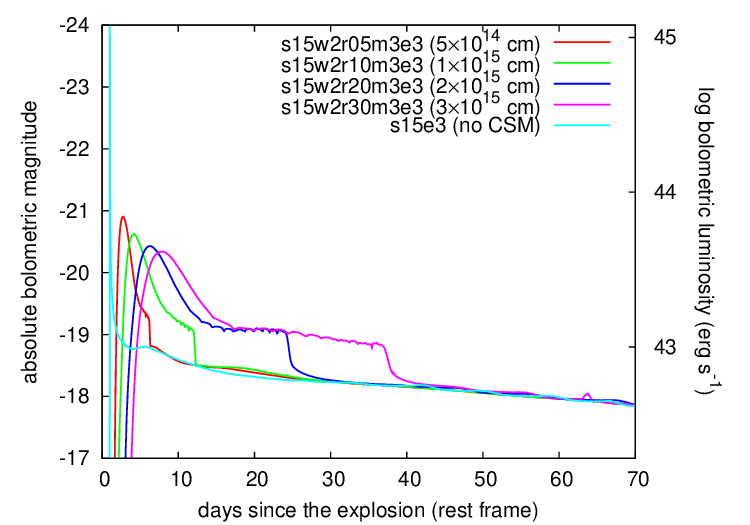}
 \caption{
Bolometric LCs of the models with different CSM radii.
The top panel shows the LCs from the models with the mass-loss rate
$10^{-2}~M_\odot~\mathrm{yr^{-1}}$ and the bottom panel
shows those with $10^{-3}~M_\odot~\mathrm{yr^{-1}}$.
The bolometric LC of the model s15e3 is also shown for comparison.
}
\label{radiibol}
\end{center}
\end{figure}

\begin{figure*}
\begin{center}
 \includegraphics[width=\columnwidth]{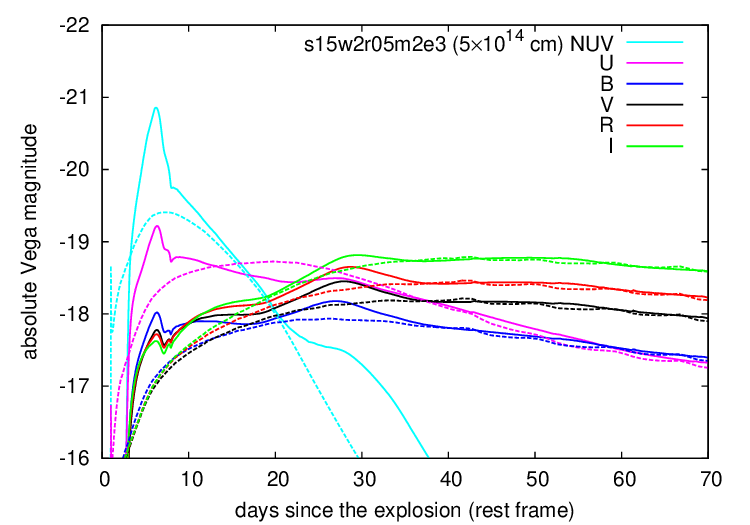}
 \includegraphics[width=\columnwidth]{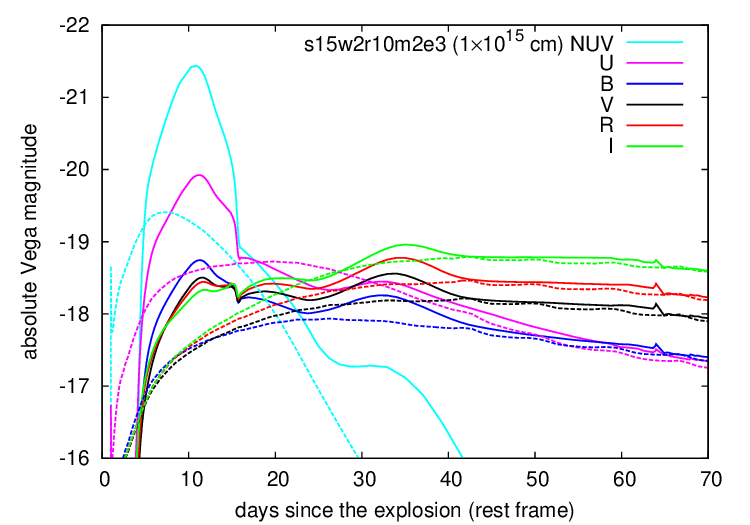}\\
 \includegraphics[width=\columnwidth]{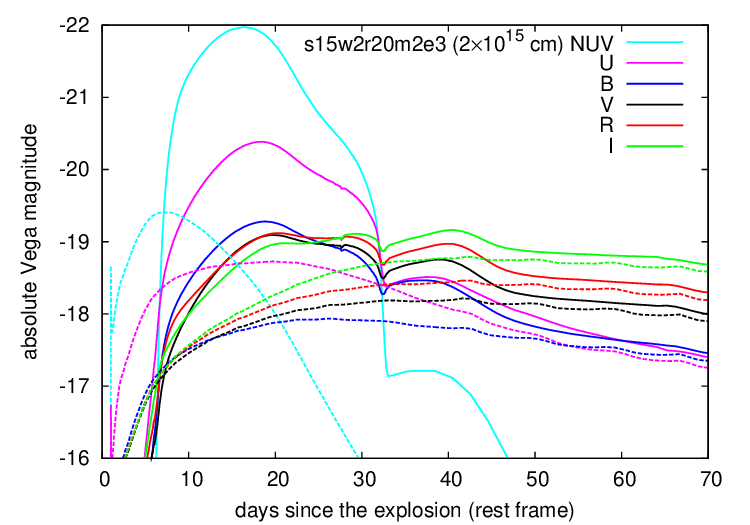}
 \includegraphics[width=\columnwidth]{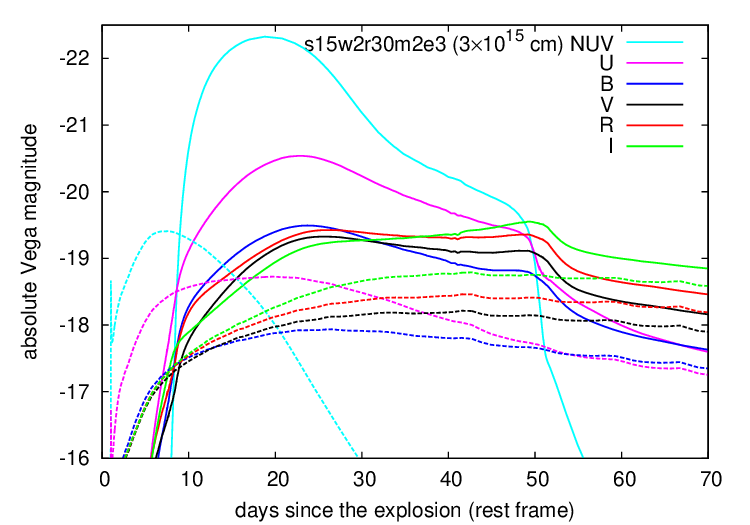}
 \caption{
Multicolor LCs of the models with different radii (solid lines).
The dashed lines are the multicolor LCs of the model s15e3
for comparison. LCs of the same color are from the same filter.
Note that the scale of $y$-axis in the right bottom panel (s15w2r30m2e3)
is different from those in the other panels.
}
\label{radiicol}
\end{center}
\end{figure*}

\subsubsection{Dependence on Radius}\label{radii}
In this section, the effect of the CSM radius on LCs is investigated.
To see the effect of the CSM radius, we fix the following parameters;
the explosion energy ($3\times10^{51}$ erg),
the density slope ($\alpha=2$), and
the progenitor inside (s15).
We try two mass-loss rates, $10^{-2}~M_\odot~\mathrm{yr^{-1}}$ and
$10^{-3}~M_\odot~\mathrm{yr^{-1}}$.
We adopt four CSM radii to see the effect, i.e.,
$5\times 10^{14}$ cm (s15w2r05m2 and s15w2r05m3),
$1\times10^{15}$ cm (s15w2r10m2 and s15w2r10m3),
$2\times 10^{15}$ cm (s15w2r20m2 and s15w2r20m3), and
$3\times10^{15}$ cm (s15w2r30m2 and s15w2r30m3).
With the constant CSM velocity $10^{6}~\mathrm{cm~s^{-1}}$,
the mass loss in the models lasts
16 years, 32 years, 64 years, and 96 years, respectively.
The density structures of the pre-SN models are shown
in Figure \ref{radiidensity}.

Figure \ref{radiibol} shows the bolometric LCs of the models with
different CSM radii.
The durations of the round phases in the IPP
are longer for the models with the larger CSM radius.
This is because the diffusion times of the LCs are longer for the models
with the larger CSM radius.
On the other hand, the maximum luminosities of the round phases
decrease as the CSM radius increases.
As the explosion energy and the density structure are similar in each
model, the radiation energy released by the shock interaction is also
close to each other (see Table \ref{wcsmlist} for the radiation energy
emitted in each model).
Therefore, the difference in the maximum luminosities is caused by the
difference in the diffusion timescales.
Even if the same energy is released in the
same timescale, photons are more scattered and distribute
more uniformly in CSM for models with longer diffusion timescales.
Thus, the luminosity, i.e., radiation energy released from the CSM surface
in a unit time, becomes lower for the models with larger CSM radii.
The flat phase of IPP is also longer for the models with the larger radius.

\begin{figure}
\begin{center}
 \includegraphics[width=\columnwidth]{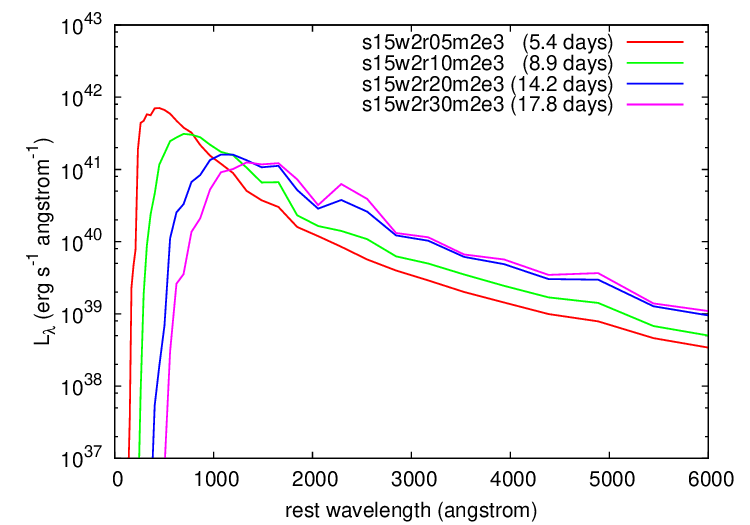}
 \caption{
SEDs of the models with different radii 
 at the epoch of the maximum 
bolometric luminosity.
The mass-loss rate of the models is $10^{-2}~M_\odot~\mathrm{yr^{-1}}$.
}
\label{radiised}
\end{center}
\end{figure}

The multicolor LCs of the models with $10^{-2}~M_\odot~\mathrm{yr^{-1}}$
are shown in Figure \ref{radiicol}.
Although s15w2r05m2e3 
has the brightest peak bolometric luminosity among the models,
the NUV and $UBVRI$ band LCs of the model are the faintest.
This is because the more compact CSM is, the hotter CSM becomes.
The photosphere of the model s15w2r05m2e3 is too hot during
the IPP to emit the radiation in the NUV and $UBVRI$ bands.
The SEDs of the models at the bolometric peak
are shown in Figure \ref{radiised}.

\subsubsection{Dependence on Density Slope}\label{denstruc}
The CSM density slope of $\rho\propto r^{-2}$ results
from the steady flow from the central progenitor.
However, if mass loss is not steady,
the density slope does not necessarily have the density
structure $\rho\propto r^{-2}$ and can be shallower
$(\alpha < 2)$
or steeper ($\alpha>2$).
For example, the shallower density slope of CSM is suggested
in the modeling of the spectra of Type IIn SNe \citep[e.g.,][]{94w}.
We calculate the LCs with the slope $\alpha=1.5$.
As we are fixing the CSM velocity, 
mass-loss rates should change with time to have the density slope
$\alpha=1.5$.
The CSM radius ($2\times 10^{15}$ cm), the explosion energy ($3\times 10^{51}$
erg), and the progenitor inside (s15) are fixed in this section.
To see the effect of density slopes, we calculate the models with the
same CSM masses as the models with the density slope $\alpha=2$,
i.e., 6.5 \Msun~(s15w1.5r20m1), 0.65 \Msun~(s15w1.5r20m2), 0.065 \Msun~(s15w1.5r20m3),
0.0065 \Msun~(s15w1.5r20m4).
The density structures of the pre-SN models are shown in Figure
\ref{masslossdensity}.

\begin{figure}
\begin{center}
 \includegraphics[width=\columnwidth]{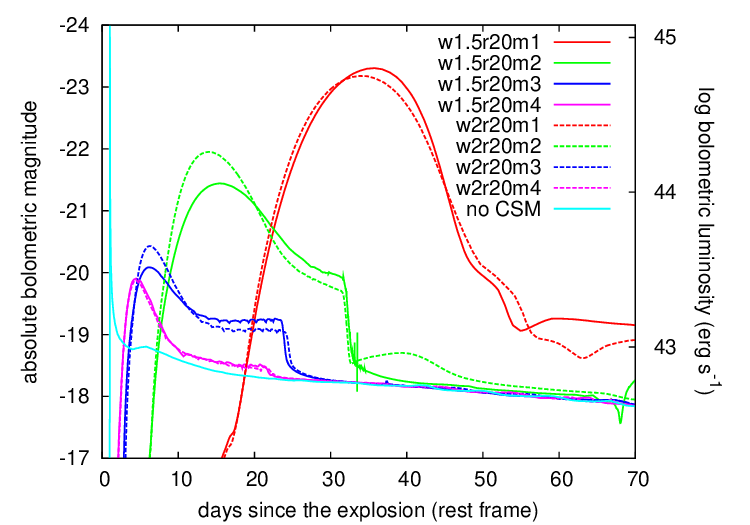}
 \caption{
Bolometric LCs of the models with different CSM density
slopes and mass-loss rates.
The bolometric LCs shown with solid lines have
CSM density slope of $\rho\propto r^{-1.5}$
and those shown with dashed lines have
CSM density slope of $\rho\propto r^{-2}$.
's15' in the model names are omitted in the figure.
The bolometric LC of the model s15e3 (no CSM) is also shown for comparison.
}
\label{densitybol}
\end{center}
\end{figure}

Figure \ref{densitybol} shows the bolometric LCs.
The dependence on CSM mass in the case $\alpha=1.5$ is similar to
that of the case $\alpha=2$.
Looking into the bolometric LCs with the same CSM mass,
the LCs with $\alpha=1.5$ are fainter until around
the bolometric peak and then become brighter.
This is because in the case of $\alpha=1.5$, CSM
is denser outside and thinner inside compared to
the case of $\alpha=2$ (Figure \ref{masslossdensity}).
Since the kinetic energy is more efficiently converted to the radiation energy
with denser CSM, LCs from shallower CSM become brighter at later epochs.
Although the luminosity of LCs is affected by density slopes,
the durations of the round phase and the epochs of
the sudden drop in the cases of $\alpha=1.5$ are similar to the cases of $\alpha=2$ and
are not strongly affected by density slopes.

The multicolor LCs are similar to those
shown in Figure \ref{masslosscol}.
The NUV absolute Vega magnitudes become as bright as 
$-22.8$ mag, $-21.7$ mag, $-20.7$ mag, and $-20.0$ mag
for the models s15w1.5r20m1e3,
s15w1.5r20m2e3, s15w1.5r20m3e3, and s15w1.5r20m4e3, respectively.

\subsubsection{Dependence on Explosion Energy}
In this section, we look into the effect of the explosion energy
on LCs.
As the higher explosion energy leads to the higher kinetic energy
of the SN ejecta,
the luminosities of the LCs during the IPP
are expected to be higher with the higher explosion energies.
The fixed parameters in this section are
the mass-loss rate ($10^{-2}~M_\odot~\mathrm{yr^{-1}}$),
the CSM radius ($2\times 10^{15}$ cm), the density slope ($\alpha=2$),
and the progenitor (s15).
The explosion energies we adopt are
$1\times10^{51}$ erg (s15w2r20m2e1),
$3\times10^{51}$ erg (s15w2r20m2e3),
$5\times10^{51}$ erg (s15w2r20m2e5), and
$7\times10^{51}$ erg (s15w2r20m2e7).
The density structure of the models in this section
is the same as that of the model
s15w2r20m2 (Figure \ref{masslossdensity}).

\begin{figure}
\begin{center}
 \includegraphics[width=\columnwidth]{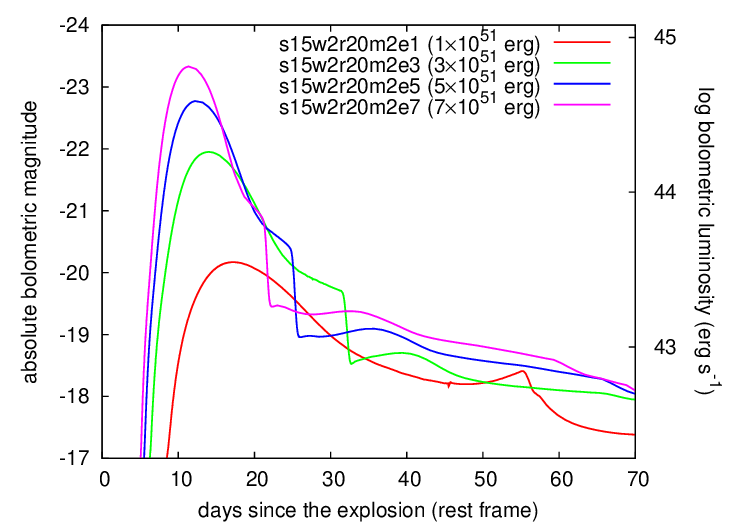}
 \caption{
Bolometric LCs from the model s15w2r20m2 with
different explosion energies.
}
\label{energybol}
\end{center}
\end{figure}

Figure \ref{energybol} is the bolometric LCs of the explosions with
different explosion energies.
As expected, the LCs become brighter with higher explosion energies.
Since the shock propagates faster in a higher energy model,
the model has a shorter rising time and a shorter duration.
However, the LCs are less sensitive to explosion energies than
mass-loss rates and radii.
This is because the CSM parameters have direct effect on diffusion timescales,
while the explosion energy determines the strength of the shock wave where
photons are emitted.

The multicolor LCs of all the models shown in this section
are similar to those of the model
s15w2r20m2e3 in Figure \ref{masslosscol} with
different rising times, durations, and brightness.
The rising times and durations are the same as those of the corresponding
bolometric LCs and
the NUV absolute Vega magnitudes become as bright as $-21.0$ mag,
$-22.4$ mag, and $-22.8$ mag
for the models s15w2r20m2e1, s15w2r20m2e5, and s15w2r20m2e7, respectively.

\subsubsection{Dependence on Progenitor}\label{progenitor}
\begin{figure}
\begin{center}
 \includegraphics[width=\columnwidth]{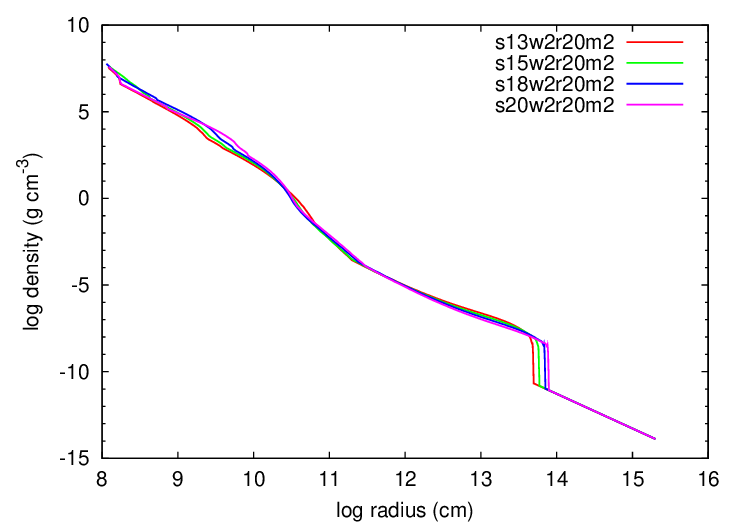}
 \caption{
Density structures of the pre-SN models with
different progenitor models inside.
}
\label{progenitordensity}
\end{center}
\end{figure}

In the previous sections, the progenitor model inside CSM is
fixed to s15. As LCs during the IPP are powered by the shock interaction
of SN ejecta and CSM,
the effect on LCs due to the difference in progenitors (RSGs) inside
is expected to be small.
To confirm this, we calculate the LCs of the models
which have different progenitors but the same CSM parameters.
The CSM parameters fixed are
the mass-loss rate ($10^{-2}~M_\odot~\mathrm{yr^{-1}}$),
the CSM radius ($2\times 10^{15}$ cm), and
the density slope ($\alpha=2$).
The explosion energy is also fixed to $3\times10^{51}$ erg.
We use Type IIP SN progenitors
s13 (s13w2r20m2),
s15 (s15w2r20m2),
s18 (s18w2r20m2),
and s20 (s20w2r20m2)
to see the effect.
Figure \ref{progenitordensity} shows the density structures
of the models. We note that the CSM density structures
of the models are slightly different from each other
because the radius where CSM is connected
to the central progenitor
is different depending on the pre-SN models.

Figure \ref{progenitorbol} shows the LCs with different progenitors. 
Roughly speaking, the LCs are similar to each other because
all the progenitors are RSGs and
properties of progenitors, like radii or density structures,
are not dramatically different from each other.
There are slight differences in the maximum luminosities
and the epochs of the sudden drop in the LCs due to
the slight difference in those properties of progenitors.

The multicolor LCs are also similar to those of the model
s15w2r20m2e3 (Figure \ref{masslosscol}).
The NUV absolute Vega magnitudes become as bright as $-22.0$ mag,
$-21.8$ mag, and $-21.9$ mag
for the models s13w2r20m2e3, s18w2r20m2e3, and s20w2r20m2e3, respectively.

\section{Discussion}\label{discussion}
In this section, we compare our LCs obtained in the previous section
to that of UV-bright Type IIP SN 2009kf
whose LC is suggested to be affected by dense CSM \citep{09kf}.
We show that the LC of SN 2009kf is actually reproduced by
the LC models with dense CSM and we get constraints on the 
state of the CSM around the progenitor of SN 2009kf at the
pre-SN stage.

\subsection{Observations of SN 2009kf and Light Curve Modeling}
SN 2009kf was discovered by Pan-STARRS 1 survey
and observed by $GALEX$ using the NUV filter \citep{09kf}.
The observations of SN 2009kf by $GALEX$ revealed
its distinguishing features:
SN 2009kf was continued to be bright in NUV for more than 10 days
and it was also bright in the optical bands during the same period.
This feature is difficult to be explained by Type IIP SN models without
dense CSM. This is because,
after the shock breakout, the UV LCs of Type IIP SNe without dense CSM
decline rapidly due to the adiabatic cooling
of the ejecta and the absorptions by iron group elements.
Therefore, the optical brightness of Type IIP SNe without dense CSM increases
as the ejecta cools down with the decreasing UV brightness (Figure \ref{s15eX}).
\citet{utrobin09kf} try to model the LC of SN 2009kf without dense CSM and they find that large explosion
energy ($2\times10^{52}~\mathrm{erg}$) is required to obtain
the LC of SN 2009kf.

One big uncertainty is in the extinction of SN 2009kf.
Although the galactic extinction is negligible
\citep[$E(B-V)=0.009$ mag;][]{galext},
the host extinction is estimated as $E(B-V)=0.32\pm0.5$ mag \citep{09kf}.
The large uncertainty in the host extinction makes it
difficult to estimate its absolute NUV magnitudes.
The redshift of the host galaxy is $0.182\pm0.002$ \citep{09kf}.

\begin{figure}
\begin{center}
 \includegraphics[width=\columnwidth]{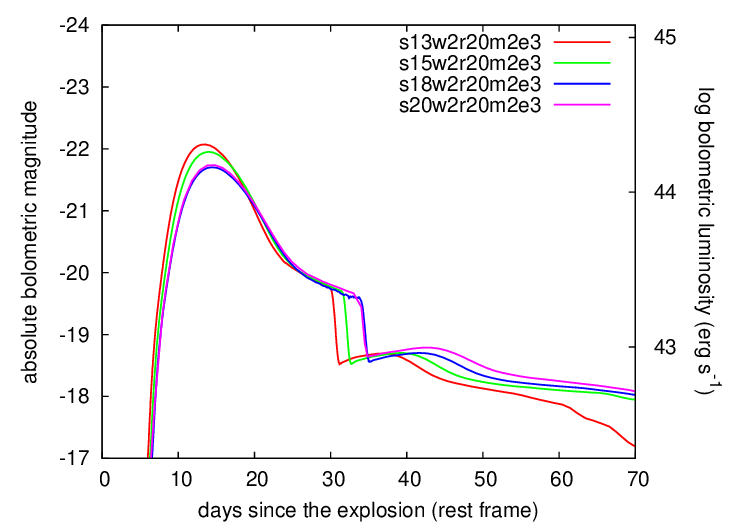}
 \caption{
Bolometric LCs of the models with different progenitors inside.
}
\label{progenitorbol}
\end{center}
\end{figure}

\begin{figure*}
\begin{center}
\includegraphics[width=\columnwidth]{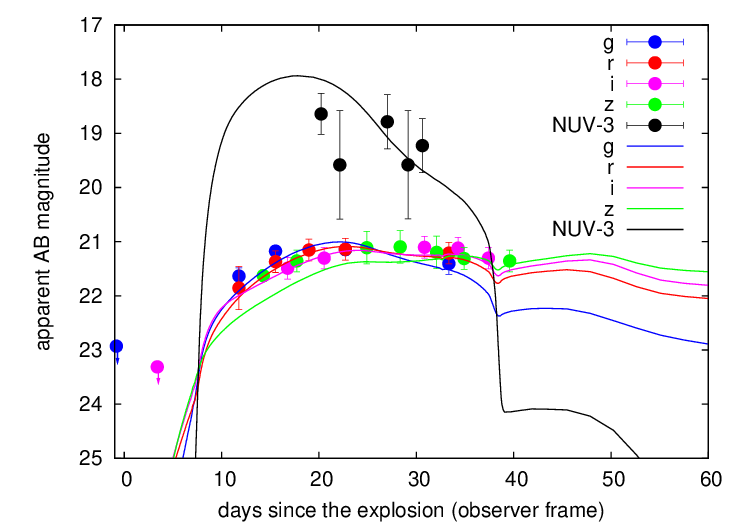}
\includegraphics[width=\columnwidth]{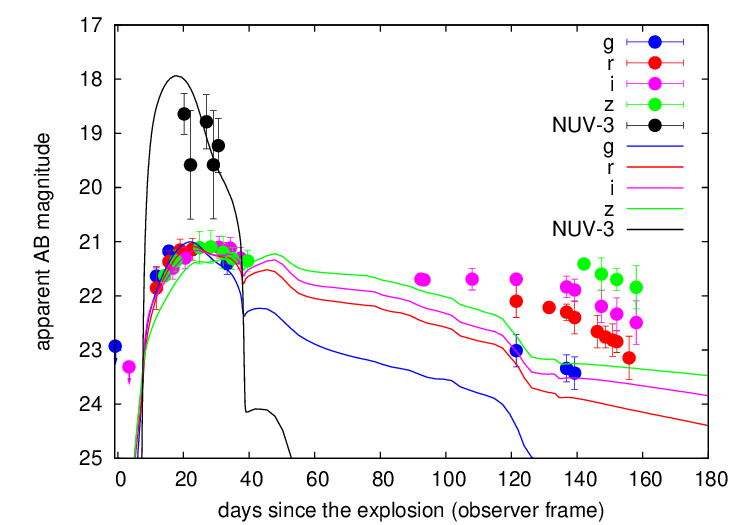}
 \caption{
Comparison of the model LCs from s15w2r20m2e3 with the observed LCs
of SN 2009kf. Each point is observed LCs and the solid lines
are the model LCs shifted to the observer frame.
The epochs of the observation points are shifted arbitrarily.
The model LCs are obtained by reddening the SEDs obtained by
numerical calculations with the host extinction
of $E(B-V)=0.22$ mag and then adopting the redshift $0.182$.
Magnitudes of NUV are shifted by 3 magnitudes in this figure.
}
\label{09kfcomp}
\end{center}
\end{figure*}

In Figure \ref{09kfcomp}, we show a comparison of the model s15w2r20m2e3
with the LCs of SN 2009kf. The SED derived at each time step
in \verb|STELLA| is reddened with the host extinction assuming
the extinction law of \citet{extinction}
and then shifted to the redshift $0.182$ by using the standard $\Lambda$CDM
cosmology with $H_0=70~\mathrm{km~s^{-1}~Mpc^{-1}}$, $\Omega_\mathrm{M}=0.3$,
and $\Omega_\mathrm{\Lambda}=0.7$.
The host extinction we apply here is $E(B-V)=0.22$ mag.
The left panel of Figure \ref{09kfcomp} shows the LCs
in the first 60 days since the explosion in the observer frame.
The solid lines are the model LCs and the points are the observations
by \citet{09kf}. The epochs of observations are shifted arbitrarily.
There is little contribution
of \Ni~produced by explosive nucleosynthesis on LCs at these epochs.
The model LCs are in good agreement with the observations.
Especially, the characteristic observational feature that the NUV LC and the optical LCs
are bright at the same epochs is well-reproduced, as well as the NUV brightness.
In the right panel of Figure \ref{09kfcomp},
the LCs at the later epochs are also shown.
Since the explosive nucleosynthesis (amount of \Ni~produced),
the progenitor model inside (mass of hydrogen layer),
and the degree of mixing of \Ni\ in H-rich layer
mainly affect the LCs at the epochs after the IPP
\citep[e.g.,][]{kaseniip,luciip,luciip2,bersten2011},
modeling this part of the LCs is out of our scope.
Especially, if we include such an effect, photosphere is expected to be located outer
than our model and photospheric velocity becomes faster.
The model shown in Figure \ref{09kfcomp}
has the explosion energy of $3\times 10^{51}$ erg
and the evolution of the photospheric velocity is shown
in Figure \ref{photov} with the observed H$\alpha$ line velocity.
Although our model LCs have slower photospheric velocity,
it is expected to increase as is mentioned above.
Large amount of \Ni~production is expected from
the long plateau phase after the IPP.
In addition, the observational facts that
the bolometric luminosity at the plateau phase is very high and
the H$\alpha$ line
velocities of SN 2009kf are very large
also indicate large explosion energy of SN 2009kf.
Note that the LCs during the IPP are not
sensitive to the central progenitor model (Section \ref{progenitor})
and progenitor models other than s15 can also work.

\subsection{CSM around the Progenitor of SN 2009kf}\label{CSMof09kf}
Given that the only observational data we are able to compare with our model
LCs are those during the IPP and that
the difference in progenitors inside (RSGs)
does not have much effect on LCs during the IPP
(Section \ref{progenitor}; Figure \ref{progenitorbol}),
it is difficult to constrain the ZAMS mass of the progenitor
of SN 2009kf.
However, as we show in Section \ref{wcsm},
the LCs of the IPP is strongly affected by the CSM parameters and
we can get constraints on them.
In this section, we try to make constraints on the CSM around the progenitor
of SN 2009kf at the pre-SN stage.

First, as discussed in Section \ref{effect},
the mass-loss rate of the progenitor should be larger than
$\sim 10^{-4}~M_\odot~\mathrm{yr^{-1}}$ to see the effect of the CSM on LCs.
The duration of the UV-bright phase of SN 2009kf indicates
that the CSM radius is larger than
$\sim1\times 10^{15}$ cm (Section \ref{radii}; Figure \ref{radiibol}).
Although a model with low explosion energy and a small CSM radius
can result in LCs with a duration similar to that of SN 2009kf,
the high bolometric luminosity and the large line velocities
of SN 2009kf are difficult to be reproduced by such low-energy explosion.

The absolute NUV magnitude of the UV-bright phase
is required to make further constraints on CSM.
However, due to the uncertainty in the host extinction,
it is difficult to get the absolute NUV magnitude of SN 2009kf.
Therefore, what we can confidently conclude is that
SN 2009kf has a LC naturally explained by the CSM interaction
(long-lasting UV brightness during the period when it is also bright in optical)
and the mass-loss rate
of its progenitor should be higher than $10^{-4}~M_\odot~\mathrm{yr^{-1}}$
just before or at the time of the explosion to see such effect on the LCs,
assuming that the CSM velocity is $10^{6}~\mathrm{cm~s^{-1}}$.
The CSM radius which is larger than $\sim 1\times 10^{15}$ cm
is inferred from the duration and brightness of the NUV LC.
If the host extinction is assumed to be $E(B-V)=0.22$ mag,
the multicolor LC model of the IPP shown in Figure \ref{09kfcomp}, which
has the CSM with the mass-loss rate
$10^{-2}~M_\odot~\mathrm{yr^{-1}}$ and the radius $2\times 10^{15}$ cm,
is consistent with the UV-bright phase of SN 2009kf.
We note that the density slope of the CSM does not have much effect
on the duration of the IPP as shown in Section
\ref{denstruc} and we cannot constrain the density slope with our model.

\begin{figure*}
  \begin{center}
    \includegraphics[width=\columnwidth]{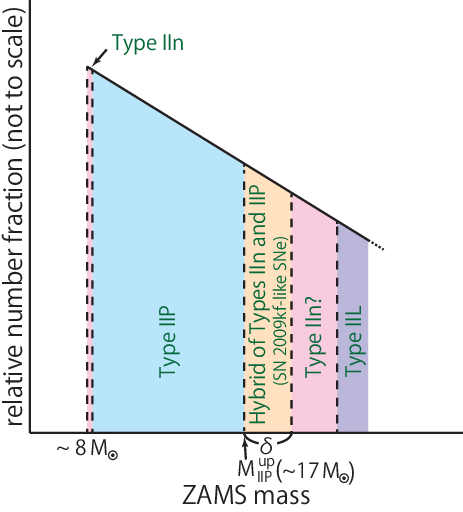}
    \includegraphics[width=\columnwidth]{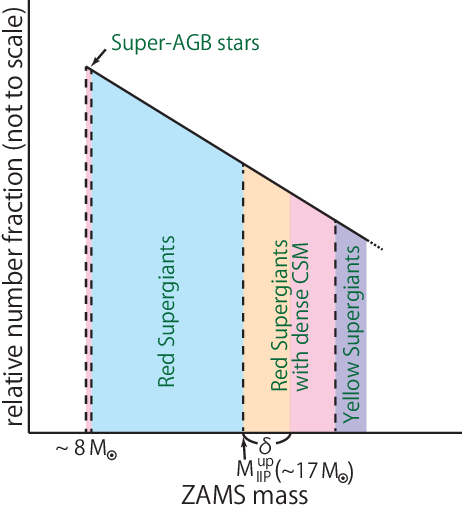}
  \end{center}
  \caption{
{\it Left:} Possible relation between ZAMS mass and SN types. 
We set the maximum ZAMS mass of Type IIP SNe as $M\mathrm{^{up}_{IIP}}$,
which is observationally suggested to be $\sim17~M_\odot$ \citep{smarttiip}.
Dense CSM may be left at the time of SN explosions at a ZAMS mass range
above $M\mathrm{^{up}_{IIP}}$. If there is enough CSM, the SNe can be Type IIn.
On the other hand, there can be a small ZAMS mass range from $M\mathrm{^{up}_{IIP}}$ to
$M\mathrm{^{up}_{IIP}}+\delta$ (with small $\delta$), in which
there is not enough CSM for the SN to continue to be Type IIn and
the SN can be a hybrid of Type IIP and Type IIn.
SN 2009kf and SN 1987C are possible candidates for such a hybrid SN.
The number fraction of SNe ($y$-axis) is determined by an initial mass function.
{\it Right:} The same as left but plotted with progenitor stars.
The explosion of RSGs without dense CSM is expected from $\sim 8~M_\odot$ 
to $M\mathrm{^{up}_{IIP}}$. RSGs explode within dense CSM above $M\mathrm{^{up}_{IIP}}$.
A yellow supergiant is observationally found to be the progenitor of
Type IIL SN 2009kr.
The ZAMS mass of the yellow supergiant is estimated as
$\sim 18-24~M_\odot$ \citep{09kr1}
and  $15^{+5}_{\ -4}~M_\odot$ \citep{09kr2},
which are consistent with our picture.
}
  \label{fig:hybrid}
\end{figure*}

\subsection{Progenitor of SN 2009kf and Its Extensive Mass Loss}\label{mostmassive}
The high bolometric luminosity at the plateau phase
and very high line velocities of SN 2009kf imply that SN 2009kf had a high explosion energy.
In addition, the long plateau phase indicates that the amount of \Ni~produced by the explosive nucleosynthesis is large.
Since Type IIP SNe from higher ZAMS mass progenitors
tend to be more energetic and produce more \Ni~
\citep[e.g.,][]{hamuy,utrobin},
the progenitor of SN 2009kf is indicated to be one of the 
most massive RSGs and may come from
the high mass end of RSGs.
Therefore, it is indicated that
the most massive RSGs
can have a very high mass-loss rate and that such extensive mass loss
can occur just before the explosions of the massive RSGs.

Not only LCs but also spectra can be affected by the existence of dense CSM.
During the IPP phase of SNe from RSGs with dense CSM,
photosphere is located in CSM and very narrow
P-Cygni profiles, which are similar to those of Type IIn SNe, are expected to
be observed (Figure \ref{photov}).
Then, after the IPP, their spectra shift to those of Type IIP SNe.
In other words, SN 2009kf-like SNe can be a hybrid type
of Type IIn and Type IIP (Figure \ref{fig:hybrid}).
If extensive mass loss of massive RSGs happen just before their explosions,
CSM mass and/or radius can be so small that the interaction of dense CSM
and SN ejecta ends in early epochs. Then, the corresponding SN
may be observed as a hybrid of Type IIn and Type IIP and SN 2009kf
might be classified as Type IIn if early spectra were taken.
If there is large and/or massive enough CSM or a shell exists that is created long before the
explosion due to extensive mass loss, the SNe may be purely Type IIn and may not
show the feature of Type IIP SNe (Figure \ref{fig:hybrid}).
Our prediction that the early spectra of SN 2009kf-like SNe have narrow lines
is what clearly differs from theoretical models suggested by \cite{utrobin09kf}.

While no spectra of SN 2009kf were taken during the IPP,
SN 1987C are suggested to have shown
such transition of the spectra from a Type IIn SN-like blue spectrum
with narrow hydrogen emission lines to Type IIP SN spectra \citep{87C}.
The Type IIn SN-like spectrum of SN 1987C is taken at 52 days since its discovery when
SN 1987C could have been approaching to the end of the IPP. At 79 days after the discovery,
the spectrum showed the P-Cygni profile of hydrogen lines and the line velocity of
H$\alpha$ was high ($6800~\mathrm{km~s^{-1}}$).
Although SN 1987C was not observed by UV, this transition of the spectra could indicate
that SN 1987C may be another sample of an explosion of a RSG within a dense CSM. 
The high H$\alpha$ line velocities indicates that
the progenitor of SN 1987C was a massive RSG.
Such early observations of SN spectra are important to
find other candidate SNe which are the hybrid of Type IIn and Type IIP.

Although we show a possible evidence for 
the existence of extensive mass loss of a massive RSG,
the mechanism is still uncertain.
A candidate for the mechanism to achieve this kind of mass loss
is the pulsations of RSGs \citep{yoonRSG,hegerRSG,Li1994}.
As shown by \citet{yoonRSG}, this mechanism can potentially
make a mass-loss rate of a RSG as high as $10^{-2}~M_\odot~\mathrm{yr^{-1}}$.
What is more, for the RSGs with ZAMS masses
around 17 \Msun, the pulsations just before the explosion and the CSM
can potentially remain dense until the time of the explosion.
Coincidentally, the minimum ZAMS mass to cause the pulsations
obtained by \citet{yoonRSG} is roughly the same as the maximum
ZAMS mass of Type IIP SN progenitors indicated by observations
\citep[$17~M_\odot$,][]{smarttiip}.
Such extensive mass loss in
massive Type IIP SN progenitors may
suppress the upper limit on the ZAMS mass
of Type IIP SN progenitors
because such mass loss can take the whole hydrogen layer
out of the progenitors.
Mass loss due to nuclear flash may also be a driving force
of the extensive mass loss \citep{neflash,lucmassloss}.

Rareness of UV-bright Type IIP SNe similar to SN 2009kf can be interpreted as
a relatively small ZAMS mass range of this event.
For example, no Type IIP SNe observed with {\it Swift} satellite
show the long term UV-brightening which is expected by CSM interaction \citep{brown}.
The rareness of UV-bright Type IIP SNe also support
that they may come from the high mass end of Type IIP SN progenitors
(Figure \ref{fig:hybrid}).
Note, however, that the rareness can also be interpreted in different ways.
For example, it is possible that RSGs generally
have extensive mass loss (due to, e.g., nuclear flashes)
but it usually occurs long before their
explosions and the mass range of the progenitors which experience
extensive mass loss just before 
the explosions is small.
Future early spectral and UV observations of Type IIP SNe are required to
get more samples of SN 2009kf-like SNe 
so that we can make constraint on the mass range of
SN 2009kf-like SNe and on the driving force of such extensive mass loss.
Observations of nebular phase spectra are also important for determining the ZAMS mass of
such SNe \citep[e.g.,][]{luciip}.

\section{Conclusions}\label{conclusions}
\begin{enumerate}
  \item   We show that the existence of dense CSM affects LCs of 
  explosions of RSGs. 
  This is because (1) shock breakout signals are elongated by
  CSM and (2) SN ejecta is decelerated by CSM.
  In particular, because of the deceleration, kinetic energy of SN ejecta is converted to
  thermal energy which is emitted as radiation and SNe can be brighter than usual.
  The LC becomes bright in UV as well as in optical.
  In addition, the photospheric velocity of early epochs is very low because the photosphere
  is located in CSM at early epochs epochs.
  The most influential
  parameters of CSM are mass-loss rates and
  radii. The mass-loss rate should be higher than $\sim 10^{-4}~M_\odot~\mathrm{yr^{-1}}$
  to show the effect of CSM.
  Higher mass-loss rates and/or larger radii lead to longer
  diffusion timescales of CSM and thus, longer durations and rising times
  of the LCs powered by the interaction.
  Density slopes and explosion energies also slightly change LCs.
  The difference in Type IIP SN progenitor (RSG) models inside CSM are not so sensitive
  to LCs. (Section \ref{LCs})

  \item The LCs of Type IIP SN 2009kf, which were bright in UV as well
	as in optical
  in early phases \citep{09kf}, can be explained by the pre-SN models
  with dense CSM. The mass-loss rate of the progenitor of
  SN 2009kf should be higher than
  $10^{-4}~M_\odot~\mathrm{yr^{-1}}$.
  The CSM radius is expected to be larger than $\sim 1\times10^{15}$ cm.
  The explosion energy of SN 2009kf is likely to be
  very high because of its high bolometric luminosity
  at the plateau phase and its high line velocities.
  The long duration of the plateau phase of SN 2009kf implies that the large amount of \Ni~
  is produced by the explosion.

  \item The high explosion energy and the large amount of \Ni~produced indicate that
  the progenitor of SN 2009kf is a massive RSG.
  Our results show that massive RSGs
  are likely to experience extensive mass loss
  exceeding $10^{-4}~M_\odot~\mathrm{yr^{-1}}$
  just before their explosions.
  The explosions of such massive RSGs
   with extensive mass loss will be SN 2009kf-like
  SNe. Their spectra show the transition
  from Type IIn SN-like spectra to Type IIP SN spectra which is observed
  in SN 1987C \citep{87C}.
  The existence of such SNe indicates that there is some mechanism to enhance
  mass loss of massive RSGs, like pulsations discussed by
  \citet{yoonRSG} or nuclear flashes suggested by \citet{neflash}.
  Such a mechanism may reduce the maximum
  ZAMS mass of Type IIP SN progenitors predicted
  by single star evolution modeling
  \citep[$\sim25M_\odot$; e.g.,][]{heger}
  as low as the observationally implicated value
  \citep[$\sim17M_\odot$; e.g.,][]{smarttiip} (Figure \ref{fig:hybrid}).

  \item Future early spectral and
  UV observations of SNe will find other SNe similar to SN 2009kf
  and provide LCs with long time coverage and spectra at the IPP.
  Large samples and detailed observations of SN 2009kf-like SNe
  can reveal the mass-loss mechanism of RSGs 
  as well as the nature of SN 2009kf-like SNe.

\end{enumerate}

\section*{Acknowledgments}
We thank the referee for his/her comments which improved the text very much.
We also thank Christopher S. Kochanek for his comment.
All the numerical calculations were carried out on the general-purpose
PC farm at Center for Computational Astrophysics, CfCA, of National
Astronomical Observatory of Japan.
T.M. would like to thank Naoki Yoshida and Keiichi Maeda for
their supports for his visits to University of Texas at Austin
and Max Planck Institute for Astrophysics.
The work of S.B., P.B., and E.S. in Russia is supported partly by the
grants RFBR 10-02-00249, 10-02-01398
by "Scientific School"  grants 3458.2010.2, 3899.2010.2,
by contract with Agency for Science and Innovation No.~02.740.11.0250,
SNSF grant  No.~IZ73Z0-128180/1 under the program SCOPES.
This research is supported in part
by a grant from the Hayakawa Satio Fund awarded by the Astronomical
Society of Japan. 
This research has also been supported in part by World Premier International
Research Center Initiative, MEXT, Japan.

\label{lastpage}

\end{document}